\documentclass[fleqn,usenatbib]{mnras}

\usepackage{newtxtext,newtxmath}

\usepackage{graphicx}	
\usepackage{amsmath}	
\usepackage{amssymb}	

\usepackage{hyperref}
\usepackage{tabu} 
\usepackage{multirow} 
\usepackage{import} 
\usepackage{gensymb} 
\title[Enhanced compact AGN emission in red QSOs]{Fundamental differences in the radio properties of red and blue quasars: enhanced compact AGN emission in red quasars}

\author[V. Fawcett et al.]
{V. A. Fawcett,$^{1}$\thanks{E-mail: victoria.fawcett@durham.ac.uk}
D. M. Alexander,$^{1}$
D. J. Rosario,$^{1}$
L. Klindt,$^{1}$
S. Fotopoulou,$^{1,2}$
\newauthor
E. Lusso,$^{3,4}$
L. K. Morabito,$^{1}$
\& G. Calistro Rivera$^{5}$
\\
$^{1}$Centre for Extragalactic Astronomy, Department of Physics, Durham University, DH1 3LE, UK\\
$^{2}$ HH Wills Physics Laboratory, University of Bristol, Tyndall Avenue, Bristol, BS8 1TL, UK\\
$^{3}$Dipartimento di Fisica e Astronomia, Universit\`a di Firenze,  via G. Sansone 1, I-50019 Sesto Fiorentino, Firenze, Italy\\
$^{4}$Osservatorio Astrofisico di Arcetri, Largo Enrico Fermi 5, I-50125 Firenze, Italy \\
$^{5}$European Southern Observatory, Karl-Schwarzchild-Strasse 2, 85748, Garching bei Mnchen, Germany
}

\pubyear{2020}

\begin{document}
\label{firstpage}
\pagerange{\pageref{firstpage}--\pageref{lastpage}}
\maketitle

\begin{abstract}
We have recently used the Faint Images of the Radio Sky at Twenty-centimeters (FIRST) survey to show that red quasars have fundamentally different radio properties to typical blue quasars: a significant (factor $\approx$\,3) enhancement in the radio-detection fraction, which arises from systems around the radio-quiet threshold with compact ($<$\,5$''$) radio morphologies. To gain greater insight into these physical differences, here we use the DR14 Sloan Digital Sky Survey (SDSS) and more sensitive, higher resolution radio data from the Very Large Array (VLA) Stripe 82 (S82) and VLA-COSMOS~3~GHz (C3GHz) surveys. With the S82 data, we perform morphological analyses at a resolution and depth three times that of the FIRST radio survey, and confirm an enhancement in radio-faint and compact red quasars over typical quasars; we now also find tentative evidence for an enhancement in red quasars with slightly extended radio structures (16--43~kpc at $z$\,$=$\,1.5). These analyses are complemented by C3GHz, which is deep enough to detect radio emission from star-formation processes. From our data we find that the radio enhancement from red quasars is due to AGN activity on compact scales ($\lesssim$\,43~kpc) for radio-intermediate--radio-quiet sources ($-$5\,$<$\,$\mathcal{R}$\,$<$\,$-$3.4, where $\mathcal{R}$\,$=$\,$L_{\textrm{1.4GHz}}/L_{\textrm{6$\upmu$m}}$), which decreases at $\mathcal{R}$\,$<$\,$-$5 as the radio emission from star-formation starts to dilute the AGN component. Overall our results argue against a simple orientation scenario and are consistent with red quasars representing a younger, earlier phase in the overall evolution of quasars.
\end{abstract}


\section{Introduction}
Quasi-stellar objects (QSOs), also known as quasars, are the most powerful class of Active Galactic Nuclei (AGN). Their extremely high bolometric luminosities (up to $10^{47-48}$~erg~s$^{-1}$) are now known to be caused by accretion onto a supermassive black hole (SMBH; $10^8$--$10^9$~M$_{\odot}$) near the Eddington limit which places them as some of the most luminous objects in the Universe. 

Due to the unobscured view of the SMBH accretion disc, which peaks in the ultra-violet (UV), the majority of Type~1 QSOs have very blue optical colours. However, there is a small but significant subset with redder optical-infrared colours (coined as ``red QSOs''). Although red QSOs have been well studied in the literature \linebreak\citep{Webster1995,Serjeant1996,kim98,rich,glik,georg,Urrutia_2009,ban12,glik12,kim18,klindt}, their exact nature remains unclear.

The origin of the red colours has been widely debated: for the majority of red QSOs the reddening appears to be due to extinction by dust e.g. \citep{Webster1995,glik,klindt}, although a red synchrotron component or stellar contamination from the host-galaxy may also contribute in some systems \citep{whiting}. However, it is unclear whether this dust is on host-galaxy or nuclear scales \citep{alex_hick}. The latter could just be a consequence of the AGN orientation model \citep{pad}, with red QSOs representing a grazing incidence viewing angle through the dusty torus. An alternative hypothesis is that red QSOs represent a rapid evolutionary phase that links obscured star-formation with AGN activity \citep{hop6,hop,alex}. In this context, merger-driven AGN activity is thought to drive a starburst phase, resulting in obscuration by dust in the early stages which is then blown out through AGN-driven outflows (commonly referred to as ``AGN feedback''), eventually resulting in an unobscured AGN (within the context of our study, a blue unobscured QSO). Some studies claim to see merger induced activity in red QSOs \citep{Urrutia_2008,urrutia12,glik15}. However, it is less clear that red QSOs systematically show an enhancement in merger signatures when compared to typical QSOs (e.g., \citealt{zak19}).

Taking a novel approach to uniformly define their QSO samples, \cite{klindt} used Sloan Digital Sky Survey (SDSS) DR7 data \citep{schneid} to optically select the top, bottom and middle 10 percentiles of the redshift-dependent observed optical colour distribution to create a red, blue and control (``typical'') QSO sample, respectively. They found a factor $\approx$\,3 larger radio detection rate in the red QSO sample making use of Faint Images of the Radio Sky at Twenty-centimeters (FIRST; \citealt{first}) data, compared to both the blue and control QSOs. The radio-detection enhancement was driven by red QSOs with compact radio morphologies ($<$\,5$''$; $<$\,43~kpc at $z$\,$=$\,1.5) and luminosities placing them around the radio-loud/radio-quiet threshold. These results rule out the simple orientation model, suggesting differences in the ``environment'' between red QSOs and typical QSOs which may be driven within the evolutionary sequence\footnote{\label{enviroment}In this work we use the term ``environment'' to indicate the environment on anything from nuclear and host-galaxy scales to much larger physical scales (i.e., the dark matter halo).}. 

QSOs exhibit a wide range of radio morphologies, with some that display large-scale jets and lobes that can extend over Mpc scales (e.g. \citealt{krish}). The main classifications can be split into \textit{compact}, where the radio emission is spatially unresolved on scales of a few arcseconds, and \textit{extended} morphologies. The extended category includes the spectacular Fanaroff-Riley (FR) type I and II \citep{fr} systems, as well as resolved diffuse emission. This binary distinction is simplistic, with sources that can show small-scale jets (few pc to tens of kpc), such as compact steep spectrum (CSS; \citealt{fanti90,fanti}) or gigahertz-peaked spectrum (GPS; \citealt{stang,odea}) sources. Recent studies have shown that FRII sources also have low-powered jets with low radio luminosities \citep{mingo}.

In this work we use deep and high-resolution radio data to further investigate the origins of the difference in the radio properties between red QSOs and typical QSOs down to kpc scales and significantly fainter radio fluxes than explored in \cite{klindt}. We use two different high-resolution Very Large Array (VLA) radio surveys: VLA Stripe~82 \citep{hodge} and VLA-COSMOS 3~GHz \citep{smol}, which are $\sim$\,3 and 38 times deeper than the FIRST survey, respectively. We select our red QSO sample from the SDSS DR14 catalogue \citep{paris} and follow a similar approach to \cite{klindt}, using the SDSS colours to define red and control QSOs (see Section~\ref{sec:sample}). In Section~\ref{sec:lum} we explore the radio enhancement in red QSOs down to lower radio fluxes than in \cite{klindt} and test how it varies across the radio-loudness plane. In Section~\ref{sec:stack} we use median stacking to explore the radio properties of the undetected population, pushing far below the survey flux threshold and in Section~\ref{sec:morph} we focus on the morphological properties of our red QSO sample, probing finer scales than in \cite{klindt}. In Section~\ref{sec:red_emis} we use the deeper COSMOS data to constrain the star-formation contribution to the radio emission from the QSOs and in Section~\ref{sec:red_emis_overall} we comment on the overall fraction of QSOs that have radio emission potentially dominated by AGN and non-AGN processes. Our results add weight to the emerging picture that red QSOs are fundamentally different to typical QSOs. Throughout our work we adopt a flat $\Lambda$-cosmology with $H_0$\,$=$\,70~kms$^{-1}$Mpc$^{-1}$, $\Omega$\textsubscript{M}\,$=$\,0.3 and $\Omega_{\Lambda}$\,$=$\,0.7.

\section{Data sets and methods}\label{sec:sample}
\begin{figure}
\begin{center}
\includegraphics[width=3.2in]{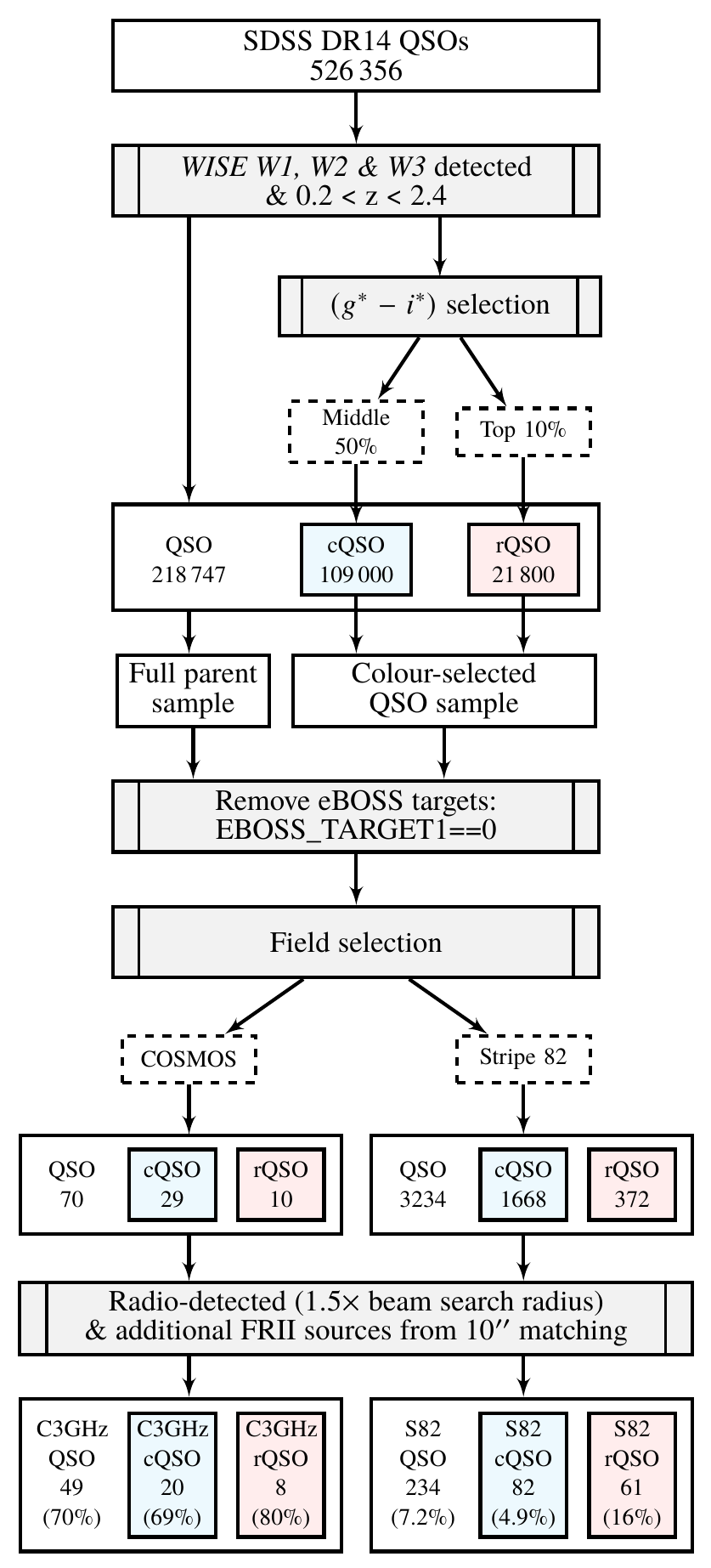}
\caption{A schematic diagram illustrating the selection process used to obtain our radio-detected samples, starting from the full SDSS DR14 quasar catalogue. We selected QSOs with redshifts 0.2\,$<$\,$z$\,$<$\,2.4 and matched to \textit{WISE} using the NASA/IPAC query engine with a 2$\farcs$7 search radius, requiring a SNR\,$>$\,2 in $W1$, $W2$ and $W3$. Our QSO colour selection was then applied to identify rQSOs and cQSOs as the top 10\% and middle 50\% of the redshift dependent observed $(g^*-i^*)$ colour distribution, respectively. We focus our study on the VLA-COSMOS 3~GHz and 1.4~GHz Stripe 82 surveys, matching the QSOs in these fields to the radio catalogues using a search radius 1.5 times the survey beam size; we additionally include FRII-like sources with weak radio cores but bright lobes that are missed with this approach using a 10$''$ search radius. Due to the discrepancy in the QSO targeting between the two regions of Stripe 82, eBOSS-targeted QSOs were removed before field selection (see Appendix~\ref{sec:bias}).}
\label{flow}
\end{center}
\end{figure}

In this paper we explore the high-resolution radio properties of SDSS optically selected QSOs at 0.2$\,<$\,$z$\,$<$\,2.4. The overall quasar selection process used is similar to that adopted in \cite{klindt}. However, in our work we now select QSOs from the SDSS DR14 Quasar Catalogue \citep{paris}, as opposed to \cite{klindt} who used the SDSS DR7 catalogue \citep{schneid}, which provides a factor $\approx 5$ improvement in sample size, as well as going almost two magnitudes deeper in the optical. In this section we describe our selection strategy, the radio surveys utilised, and the key measurements extracted from the multi-wavelength data in the SDSS. Fig.~\ref{flow} shows a schematic representation of the sample selection, which we describe in more detail in the following subsections.

\subsection{Parent sample: optical and mid-infrared data}
\subsubsection{SDSS DR14 Quasar Catalogue}\label{sec:sdss}
The SDSS DR14 Quasar Catalogue \citep{paris} contains 526\,356 spectroscopically selected QSOs with luminosities $M_i[z=$~$2]$~<~$-20.5$, out to redshifts around $z=7$. The survey covers a region of 9376~deg$^{2}$ and consists of various different targeting campaigns. The catalogue includes previous spectroscopically-confirmed QSOs from the SDSS-I and II Legacy surveys \citep{york} with QSOs targeted by the Baryon Oscillation Spectroscopic Survey (BOSS; \citealt{boss}) in SDSS-III \citep{eise} and the extended Baryon Oscillation Spectroscopic Survey (eBOSS; \citealt{eboss}) in SDSS-IV. As explained in Section~\ref{sec:colour}, we removed the eBOSS-targeted QSOs from our final samples due to differences in the source densities of the two Stripe 82 regions.

The \cite{paris} quasar catalogue provides spectroscopic redshifts based on different estimators; in this work we use the most robust of these estimates (listed as $Z$ in the catalogue). The SDSS five-band optical photometry ($ugriz$) is also utilised, corrected by the associated band-dependent Galactic extinction measurements. 

\subsubsection{Mid-infrared counterparts: matching to WISE}\label{sec:six}
We matched the SDSS DR14 quasar sample to mid-infrared (MIR) counterparts using the Wide-Field Infrared Survey Explorer (\textit{WISE}; \citealt{wise}), an all-sky survey which provides photometry in four bands (3.4, 4.6, 12 and 22~\micron). The emission from the QSO accretion disc is expected to be absorbed and reradiated by hot dust at the inner edge of the torus, leading to an infrared (IR) excess that peaks at MIR wavelengths. Therefore the MIR emission is a useful, extinction-insensitive discriminant of QSOs (e.g. \citealt{stern05,lacy05,stern12,assef}), as well as a measure of their intrinsic luminosity; for example the commonly used rest-frame 6$\upmu$m luminosity (\textit{L}\textsubscript{6$\upmu$m}). 

We used the NASA/IPAC query engine to match SDSS DR14 QSOs to the All-Sky \textit{WISE} Source Catalogue (ALL-\textit{WISE}) adopting a $2\farcs7$ search radius, which ensured a 95.5\% positional certainty \citep{lake11}, and required a detection with a signal-to-noise ratio (SNR) of greater than 2 in the \textit{WISE} \textit{W1}, \textit{W2} and \textit{W3} bands. As in \cite{klindt}, we only selected QSOs with redshifts between 0.2\,$<$\,$z$\,$<$\,2.4: the higher redshift cut ensures there is no contamination from the Lyman break in the $g^*$ band and the lower redshift cut ensures we only consider luminous QSOs in our sample.

The need for a \textit{WISE} detection in the three bands and the restriction of 0.2\,$<$\,$z$\,$<$\,2.4 reduced the number of QSOs to 218\,747, the full parent sample; see Fig.~\ref{flow}. To create the C3GHz and S82 parent samples (70 and 3234 QSOs, respectively; see Table~\ref{tab:rad_num} and Fig.~\ref{flow}), we have restricted the full parent sample to the regions covered by these two radio surveys, using the average root mean square (RMS) mosaics (see Sections~\ref{sec:s82} and \ref{sec:c3ghz}). Using \textit{WISE}, we computed the rest-frame 6$\upmu$m flux with a log-linear interpolation (or extrapolation) of the $W2$ and $W3$ bands following the approach outlined in \cite{klindt}. The flux was then converted into a luminosity (\textit{L}\textsubscript{6$\upmu$m}) using the cosmological luminosity distance.

\subsection{Radio data}
\begin{figure}
    \centering
    \includegraphics[width=3.2in]{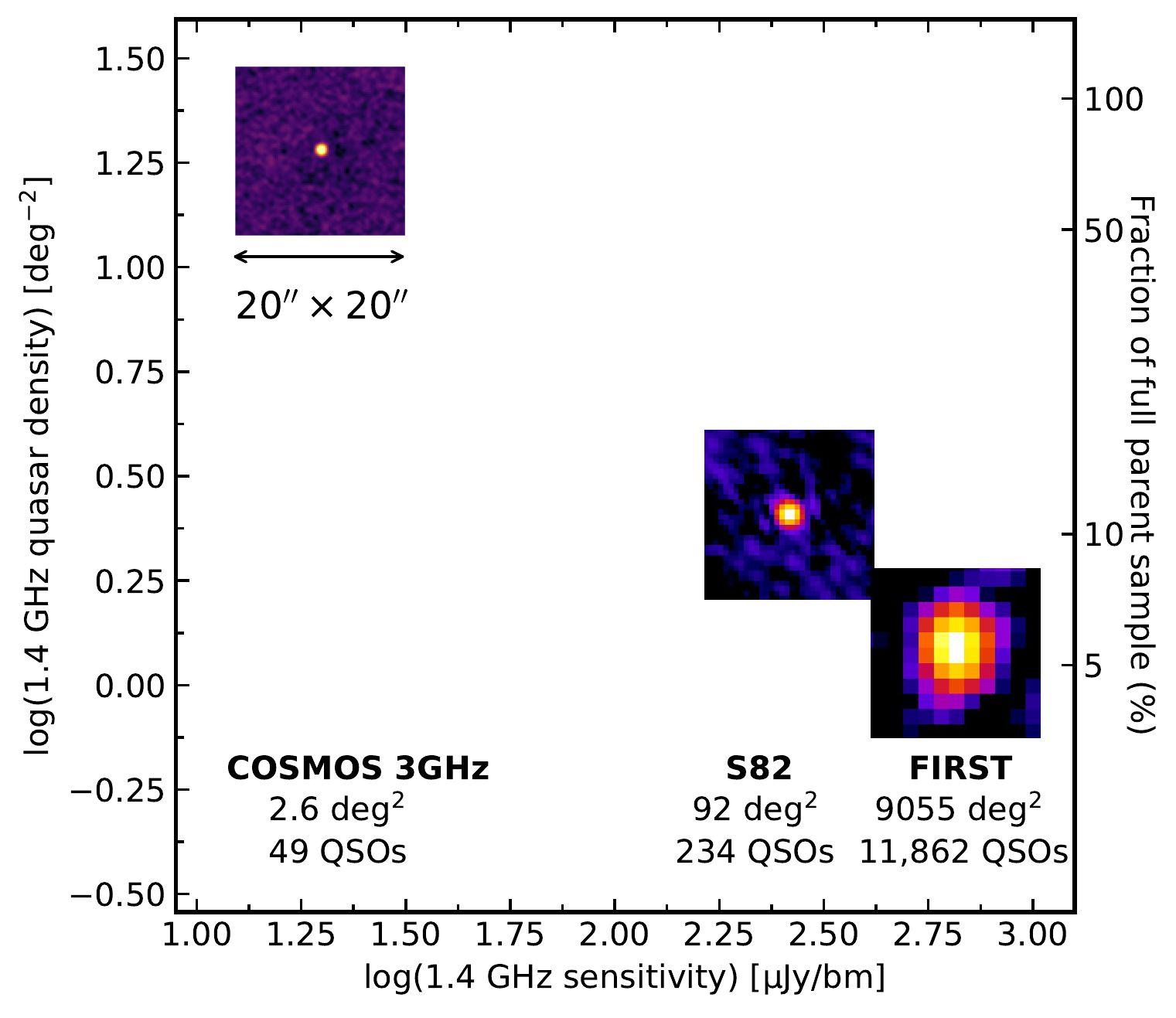}
    \caption{Radio-detected parent sample QSO source density versus $5\sigma$ 1.4~GHz sensitivity for the three radio surveys considered in this paper. For C3GHz, the 1.4~GHz sensitivity was calculated from the 3~GHz sensitivity limit assuming a uniform spectral slope of $\alpha$\,$=$\,$-$0.7. The 20$''$\,$\times$\,20$''$ thumbnails give a visual illustration of the resolution for a compact (i.e., unresolved) radio source in the different radio surveys; the number of radio-detected C3GHz, S82 and FIRST parent sample QSOs is shown underneath (see Table~\ref{tab:rad_num}).
    The source density as a fraction of the full parent sample is displayed on the right-hand axis as a reference.}
    \label{fig:survey}
\end{figure}

The main focus of this paper is to understand the differences in the radio properties between red and typical QSOs. We used data from two high spatial resolution VLA radio surveys: VLA Stripe~82 (S82; \citealt{hodge}) and VLA-COSMOS~3~GHz (C3GHz; \citealt{smol}), which we describe in more detail below. To better place these data into context with our earlier \cite{klindt} paper, we also describe the VLA FIRST survey used in that work.

Fig.~\ref{fig:survey} compares the radio source density of the C3GHz, S82 and FIRST parent sample QSOs matched to the three different radio survey catalogues, using a search radius 1.5 times the survey beam size for S82 and C3GHz, and 10$''$ for FIRST (to be consistent with \citealt{klindt}), as a function of the 5$\sigma$ sensitivity limit for each survey; 70\% of the parent QSOs are detected in C3GHz, an order of magnitude greater than in FIRST. The cutouts illustrate a compact (i.e., unresolved) radio source at the different survey resolutions.

\subsubsection{VLA Stripe 82 (S82)}\label{sec:s82}
Stripe 82 is a $\sim$\,300~deg$^2$ equatorial field that has been imaged multiple times by SDSS \citep{jiang}. It spans a right ascension of $\alpha$\,$=$\,$-$50$\degree$ to $+$59$\degree$  and a declination of $\delta$\,$=$\,$-$1.25$\degree$ to 1.25$\degree$. The high-resolution radio survey, S82 \citep{hodge}, covers $\sim$\,92~deg$^{2}$ and has a 1$\farcs$8 spatial resolution at 1.4~GHz, taken primarily in the A-configuration, to a sensitivity roughly three times below that of FIRST. We note that \cite{hey} provide a similar survey using the VLA hybrid CnB configuration, but at a lower resolution (16$''$\,$\times$\,10$''$) and shallower depth; this survey recovers some lost flux from sources in Stripe~82 with diffuse radio lobes, but results in a lower overall number of sources. As a key focus of our study is high spatial resolution, we have only used the \cite{hodge} radio survey. 

In the catalogue provided by \cite{hodge}, peak flux densities (\textit{F}\textsubscript{peak}) are derived by fitting an elliptical Gaussian model to the source. The fitted major axis used for physical size calculations are derived from the elliptical Gaussian model, fitted before deconvolution of the circular clean beam. Matching to the FIRST catalogue, \cite{hodge} recover over 97\% of the FIRST-detected QSOs.

\subsubsection{VLA-COSMOS 3~GHz (C3GHz)}\label{sec:c3ghz}
The C3GHz survey spans 2.6~deg$^2$ at 3~GHz, centred on the COSMOS field (RA~=~10:00:28.6, Dec~=~+02:12:21.0) at a $0\farcs75$ spatial resolution in the A+C configuration \citep{smol}. At its median depth ($\sim$\,7~$\upmu$Jy at $5\sigma$), C3GHz can detect a source that is equivalently 13 times deeper than the S82 sensitivity limit at 1.4~GHz, assuming a spectral index of $\alpha$\,$=$\,$-$0.7 ($S_{\nu}\propto \nu^{\alpha}$; $\sim$\,38 times deeper than FIRST); C3GHz is currently the largest and deepest radio continuum survey at such a high-resolution \citep{smol}. In the catalogue provided by \cite{smol}, \textit{F}\textsubscript{peak} is measured by fitting a two-dimensional parabola around the brightest pixel. Using a Monte Carlo method, they derived a source completeness of 55\% up to 20~$\upmu$Jy, which rises to 94\% above 40~$\upmu$Jy.

The COSMOS field has also been observed at 1.4~GHz over 2~deg$^{2}$ in an earlier VLA radio survey \citep{smol07}. This survey has a resolution of 1$\farcs$5\,$\times$\,1$\farcs$4 and an RMS of $\sim$\,10.5~$\upmu$Jy/bm ($\sim$\,5 times deeper than S82) and is also utilised in this paper when available, for comparison with the S82 data.

\subsubsection{VLA FIRST}
The FIRST radio survey covers 9055~deg$^2$ of the North Galactic Cap and Equatorial Strip in the SDSS region. FIRST has a 5$''$ resolution at 1.4~GHz taken primarily in the VLA B-configuration. The catalogue \citep{first} contains 946\,000 sources with a typical RMS of 0.15~mJy; 30\% of the FIRST sources have optical counterparts in the SDSS.

\subsection{Red and control QSO samples}\label{sec:gi}
\begin{figure}
    \centering
    \includegraphics[width=3in]{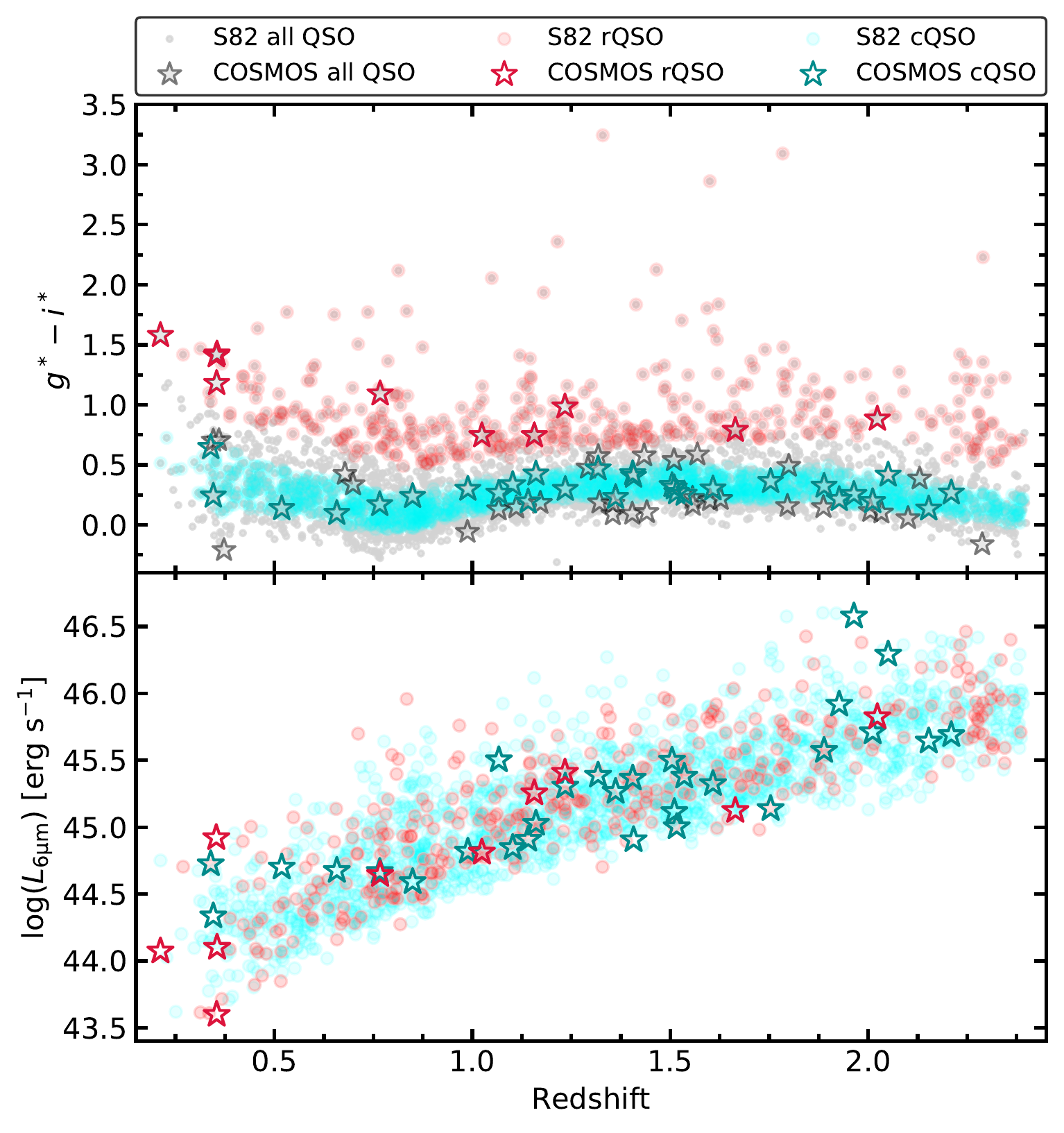}
    \caption{Sample distribution comparison of observed $g^*-i^*$ colour (top) and rest-frame \textit{L}\textsubscript{6$\upmu$m} luminosity (bottom) versus redshift. The filled circles represent S82 QSOs and the open stars represent the C3GHz QSOs: the rQSOs and cQSOs are plotted in red and cyan, respectively. The grey points in the top plot illustrate the colours of the parent sample DR14-\textit{WISE} detected QSOs within the two fields, with the eBOSS-selected sources removed (Stripe~82: 3234; COSMOS: 70). The top plot shows our redshift dependent colour-selection approach and the bottom plot shows that the range in \textit{L}\textsubscript{6$\upmu$m} and redshift for the two samples are consistent.}
    \label{fig:comp}
\end{figure}

One of the challenges in comparing between different red QSO studies is the variety of selection methods used to obtain a red QSO sample. In this paper we take the same approach as \cite{klindt}, distinguishing between red and typical QSOs based on SDSS optical photometry in a redshift dependent manner. Selecting a control sample in a consistent way allows us to robustly identify differences with respect to the red QSO population, in contrast to most other studies which use a separate control sample (often with a different selection approach and wavebands) for their comparison. Our selection is quantitatively different to that in \cite{klindt}, who defined their control QSOs as the middle 10th percentile of the $g^*-i^*$ distribution, in addition to a blue QSO sample (the bottom 10th percentile). We do not define a blue QSO sample here since \cite{klindt} demonstrated that there are no significant differences between the blue and control QSO samples and consequently we also defined a broader control QSO sample to improve the source statistics in our comparisons with the red QSOs. Our companion study by \cite{rosario} also employs a similar colour selection.

We note that by only using optical photometry we expect our colour selection to result in red QSOs with extinction values of $A_V$\,$\sim$\,0.1--0.5~mag (see Section~2.2.3 in \citealt{klindt}), missing the reddest systems (e.g. \citealt{glik12,ban} also exploit near-IR photometry to select QSOs with even redder colours); however, we highlight that a key benefit of our approach is the consistent selection of both red and control QSOs.

\subsubsection{Colour-selected samples}\label{sec:colour}
To define our sample of colour-selected QSOs we used the $g^*$ (4770~\AA) and $i^*$ (7625~\AA) band extinction-corrected photometry. Our red and control QSOs (hereafter, rQSOs and cQSOs) were selected above the top 90th percentile and within the middle 50th percentile of the observed SDSS $g^*-i^*$ distribution, respectively; the cQSOs therefore represent typical QSOs. In order to produce a redshift sensitive colour sample, the QSOs were sorted by redshift and the $g^*-i^*$ distribution was binned using contiguous redshift bins consisting of 1000 sources. This produced a sample of 130\,800 QSOs (21\,800 rQSOs and 109\,000 cQSOs; see Fig.~\ref{flow}). The $g^*-i^*$ colour distribution of the colour-selected samples is shown in the top panel of Fig.~\ref{fig:comp}. 

The SDSS-IV eBOSS survey (see Section~\ref{sec:sdss}) only covered the Eastern part of the Stripe~82 (known as East: RA\,$\lesssim$\,36~deg), which leads to a difference in the sky densities of QSOs with respect to the Western part (know as West: RA\,$\gtrsim$\,330~deg), as well as a difference in the QSO redshift distributions. Therefore, we excluded the eBOSS-targeted QSOs from the colour-selected sample and S82 parent sample (see Fig.~\ref{flow}; we note there are no eBOSS-targeted QSOs in COSMOS). This meant the two regions of Stripe~82 can be combined in a consistent way (see Appendix~\ref{sec:bias} for more details). 

For our study here, the colour-selected QSO sample (with eBOSS-targeted QSOs removed) was then restricted to the S82 and C3GHz survey regions. For S82, we defined the areas using the online catalogue matching service provided by \cite{hodge}. This provided non-zero local RMS values for detected and undetected QSOs within the survey region, which resulted in 2040 colour-selected QSOs within S82 (1668 cQSOs and 372 rQSOs). For C3GHz the RMS mosaic was used to only select QSOs within the radio-observed region of COSMOS, which resulted in 39 colour-selected QSOs (29 cQSOs and 10 rQSOs)\footnote{The VLA-COSMOS~3GHz RMS mosaic is available online \linebreak (http://jvla-cosmos.phy.hr/Home.html).}. Selecting the colours before restricting to the two fields ensured that the two colour-selected samples are representative of the entire DR14 QSO population.

The bottom panel of Fig.~\ref{fig:comp} displays the \textit{L}\textsubscript{6$\upmu$m}--$z$ distributions for the colour-selected QSO sample in the S82 and C3GHz regions. Applying the two-sided Kolmogorov-Smirnov test for the S82 and C3GHz rQSO and cQSO redshift distributions separately, we cannot rule out that the two samples are drawn from the same parent distribution at a $\sim$\,20\% and 10\% significance, respectively, which shows that the redshift distribution of the red and control samples are broadly consistent with each other.

\subsubsection{Final radio-detected samples: S82 and C3GHz}\label{sec:radio}
\begin{figure*}
    \centering
    \includegraphics[width=2.05in]{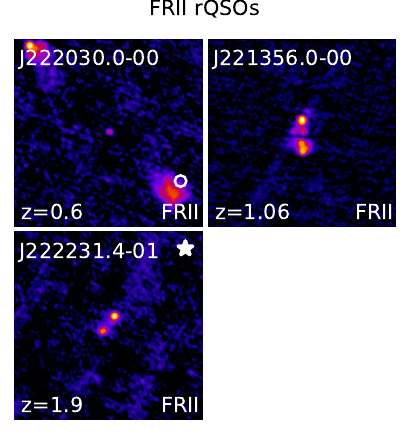}
    \hspace{1mm}
    \includegraphics[width=4in]{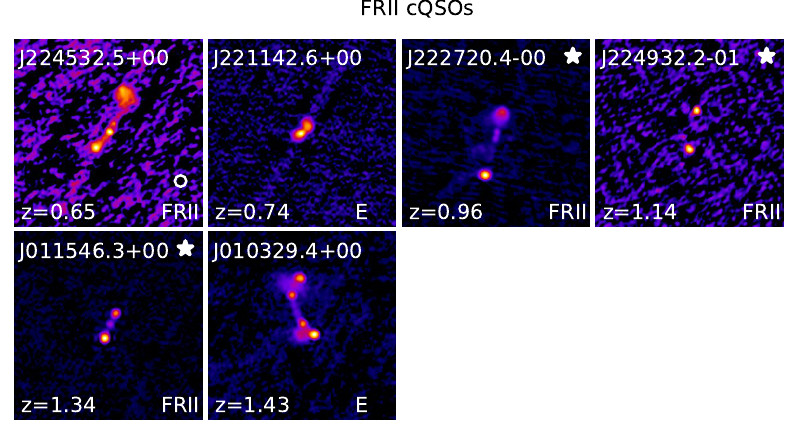}
    \caption{Thumbnails ($1\farcm5\times1\farcm5$) using data from \protect\cite{hodge} of S82 FRII-like rQSOs (left) and cQSOs (right). The morphology classification (see Section~\ref{sec:morph}) from visually inspecting the FIRST VLA data is shown in the bottom right of the image; FRII (FRII-like) or E (extended). The redshift is shown in the bottom left corner and the VLA source name is shown at the top left corner. The star in the top right corner indicates the 4 FRII sources included from the wider 10$''$ search radius. The white circle in the first thumbnail illustrates the 1$\farcs$8 beam size of the VLA data.}
    \label{fig:fr2}
\end{figure*}

\begin{table}
    \centering
    \caption{The number of parent QSOs, rQSOs and cQSOs, and their radio-detection fraction, for our two samples; see Fig.~\ref{flow} for selection process and sample numbers. The parent QSOs are selected within the two regions before the colour-selection. The radio-detection enhancement of the rQSOs in comparison to the cQSOs is also shown, where we see a clear enhancement for the S82 sample but only a tentative enhancement in the deeper C3GHz sample.}
    \begin{tabular}{p{0.85cm} p{0.8cm} |p{0.65cm} p{0.95cm} p{1.1cm} p{1.5cm}}
        \hline
        \hline
        \multicolumn{2}{c|}{\multirow{2}{*}{Sample}} & \multirow{2}{*}{Num.} & Radio & Detection & rQSO radio \\
        & & &  detected & \% & enhancement \\
         \hline
         \multirow{3}{*}{S82} & Parent & 3234 & 234  & 7.2 & - \\
         & rQSOs & 372 & 61  & 16 & \multirow{2}{*}{3.3$^{+0.6}_{-0.5}$}\\
         & cQSOs & 1668 & 82  & 4.9 & \\
         \hline
         \multirow{3}{*}{C3GHz} & Parent & 70 & 49 & 70 & - \\
         & rQSOs & 10 & 8 & 80  & \multirow{2}{*}{1.2$^{+0.2}_{-0.3}$}\\
         & cQSOs & 29 & 20 & 69  & \\
         \hline
         \hline
    \end{tabular}
    \label{tab:rad_num}
\end{table}

The colour-selected QSO samples were matched to the high-resolution radio catalogues using a search radius of 1.5 times that of the beam size (see Sections~\ref{sec:s82} and \ref{sec:c3ghz}). This resulted in an initial sample of 139 S82-detected QSOs (60 rQSOs and 79 cQSOs) and 28 C3GHz-detected QSOs (8 rQSOs and 20 cQSOs). Based on the analysis of \cite{lu} (applied to the deeper depths of these surveys), we expect a false association rate of $\sim$\,0.2\% and 0.6\% for C3GHz and S82, respectively. To take into account rarer FRII-like sources, whose cores may be too faint to be detected and whose radio lobes can extend beyond our search radius, we matched again using a 10$''$ search radius and summed the radio flux from all sources within that radius. This produced an extra 12 potential matches for the S82 data, and 1 for C3GHz. These sources were visually inspected and only the FRII-like sources were added to our samples (S82: 4; C3GHz: 0, see Fig.~\ref{fig:fr2}), with the remaining sources discounted as spurious matches (see Fig.~\ref{fig:spurious}). This gave a final sample of 143 S82-detected QSOs (61 rQSOs and 82 cQSOs) and 28 C3GHz-detected QSOs (8 rQSOs and 20 cQSOs); see Fig.~\ref{flow}. The C3GHz and S82 parent samples (70 and 3234 QSOs, respectively) were also matched to the respective radio surveys following the same method as the colour-selected samples, resulting in 49 and 234 radio-detected C3GHz and S82 parent sample QSOs, respectively. Table~\ref{tab:rad_num} displays the number of QSOs and radio-detection fraction for the parent QSOs, rQSOs and cQSOs in the two surveys, and the overall selection process is shown in Fig.~\ref{flow}.

\begin{table}
    \caption{Percentage of our colour-selected and radio-detected final samples that lie outside the AGN wedge ($W1-W2$ vs $W2-W3$) from \protect\cite{mateos}. We find that the majority of QSOs in our two samples have \textit{WISE} colours consistent with AGN.}
    \begin{center}
    \begin{tabu} to 0.45\textwidth{X|XX|XX}
        \hline
        \hline
           \multicolumn{5}{c}{Outside AGN Wedge} \\
          \hline
          & \multicolumn{2}{c|}{Stripe 82} & \multicolumn{2}{c}{COSMOS} \\
          & \centering Num. & \centering\% &\centering Num. & \centering\% \\
         \hline
          \multicolumn{5}{c}{Colour-selected} \\
         \hline
         rQSO & \centering37/372 & \centering9.9\% & \centering1/10 & \centering10\% \\
         cQSO & \centering117/1668 & \centering7.0\% & \centering1/29 & \centering3.4\% \\
         \hline
          \multicolumn{5}{c}{Radio-detected} \\
         \hline
         rQSO & \centering4/61 & \centering6.6\% & \centering0/8 &  \centering0\% \\
         cQSO & \centering6/82 & \centering7.3\% & \centering0/20 & \centering0\% \\
         \hline
         \hline
    \end{tabu}
    \end{center}
    
    \label{tab:wedge}
\end{table}

To verify that the MIR emission of our final samples is dominated by the AGN, we explored the \textit{WISE} $W1-W2$ versus $W2-W3$ colour-colour space, in which \cite{mateos} defined a region used to reliably select luminous AGN (the ``AGN wedge''). The percentage of our colour-selected and radio-detected samples that lie outside the AGN wedge are given in Table. \ref{tab:wedge}. This shows that the majority of the QSOs ($\sim$\,90--100\% for COSMOS and $\sim$\,90--95\% for Stripe~82) lie within the wedge. The outliers are predominately the low luminosity sources at all redshifts, with the majority at either the lowest or highest end of our redshift range. The low luminosity end may have a significant level of host-galaxy contamination, as also suggested in \cite{klindt}. For the high redshift end ($z>$\,2), it is known that the AGN wedge becomes less reliable at selecting AGN (see Figure 5 in \citealt{mateos}).

The rest-frame 1.4~GHz luminosities were calculated for S82 by assuming a uniform spectral slope of $\alpha$\,$=$\,$-$0.7, following \cite{alex2003}\footnote{In this paper we define the radio spectral slope ($\alpha$) as $S_{\nu}\propto \nu^{\alpha}$.}. For the C3GHz sample, the 1.4~GHz fluxes were taken from the VLA-COSMOS~1.4~GHz survey \citep{smol07} when available, and calculated from the 3~GHz fluxes assuming a uniform spectral slope of $\alpha$\,$=$\,$-$0.7 otherwise. If 1.4~GHz fluxes were available, the rest-frame 1.4~GHz luminosities were calculated using the measured spectral slope between 1.4 and 3~GHz. 

\subsection{Visual assessment of radio morphologies}\label{sec:morph_meth}
To determine the radio morphologies of the colour-selected QSOs we took the same approach as that used in  \cite{klindt}. We created $1\farcm5$ square cutouts for all sources, and visually inspected them in random order, blind to whether they were part of the rQSO or cQSO sub-samples so as not to bias our judgement. For the S82 analysis, the FIRST images were also inspected to obtain morphologies using both the S82 data and the FIRST data, which could then be used to explore the change in morphology when using higher resolution and deeper data (see Appendix~\ref{sec:first}). Here we use similar morphology classes to \cite{klindt}, where a source can be defined as compact (unresolved; i.e., point-like), extended (single sources with extended emission), FRII-like (double lobed systems with \textit{F}\textsubscript{peak,lobe}~$>$~\textit{F}\textsubscript{peak,core}) or faint; \textit{F}\textsubscript{peak}~$<$~15$\sigma$, where $\sigma$ is the typical RMS of the respective survey (S82: \textit{F}\textsubscript{peak}~$<$~1~mJy; FIRST: \textit{F}\textsubscript{peak}~$<$~3~mJy; C3GHz: \textit{F}\textsubscript{peak}~$<$~35~$\upmu$Jy). In our study, classic FRI sources (\textit{F}\textsubscript{peak,lobe}~$<$~\textit{F}\textsubscript{peak,core}) are included within the extended category and the compact-FRII class adopted in \cite{klindt} is combined within the FRII-like category due to the rarity of these systems. 

\subsection{Median stacking procedure}\label{sec:median}
Stacking is a method widely used to analyse the average properties of source populations that lie below the detection limit, with most previous studies exploring radio stacking using the FIRST radio data \citep{white06,hodge08,hodge09,Kratzer_2015}. We undertook a stacking analysis of the S82 and C3GHz radio data to probe down to $\upmu$Jy fluxes to assess the average radio properties of red and control QSOs.

We started with a sample of 1897 S82 radio-undetected QSOs (311 rQSOs, 1586 cQSOs) and 11 C3GHz radio-undetected QSOs (2 rQSOs, 9 cQSOs). After removing images with only partial radio coverage, we median stacked a final sample of 290 and 1532 S82-undetected rQSOs and cQSOs, respectively, and 2 and 9 C3GHz-undetected rQSOs and cQSOs, respectively. The samples were also stacked in four contiguous redshift bins. Peak flux measurements were obtained by fitting a two-dimensional Gaussian to the stacked map and taking its amplitude. Errors were calculated from the standard deviation of the stacked map after masking a central circular region of 10$''$; we note that the obtained flux values have not been corrected for any CLEAN or snapshot bias and so will contain additional errors of $\sim$\,10\% (see \citealt{white06}).
\section{Results}\label{sec:result}
The deep and high-resolution radio data in the Stripe~82 and COSMOS fields (see Sections~\ref{sec:s82} \& \ref{sec:c3ghz}) allows us to search for fundamental differences between red and typical QSOs. In \cite{klindt} we found that the main differences between the radio properties of red and typical QSOs occurred towards the radio-quiet end of the population, therefore our samples can be used to further probe this faint end, providing insight into the origin and scale of the radio emission. 

In Sections~\ref{sec:lum} and \ref{sec:stack} we analyse the radio-faint population; exploring the radio-loudness parameter and stacking the radio-undetected populations. In Section~\ref{sec:morph} we compare the radio morphologies of the rQSOs and cQSOs below the FIRST resolution limit, probing smaller scale radio emission.

\subsection{Radio enhancement in red QSOs}\label{sec:lum}
Comparing the radio-detection fraction of rQSOs and cQSOs, we find an overall enhancement in the radio-detection fraction for rQSOs compared to cQSOs of $\approx$\,3.3 for the S82 sample (see Table~\ref{tab:rad_num})\footnote{The radio-detection enhancement is calculated by dividing the radio-detection fraction for the rQSOs by that of the cQSOs.}. For C3GHz, within the uncertainties, there is no enhancement which is primarily due to the majority of the QSO population now being detected in the ultra-deep radio data (69\% for cQSOs and 80\% for rQSOs). 

\begin{figure*}
    \centering
    \includegraphics[width=6in]{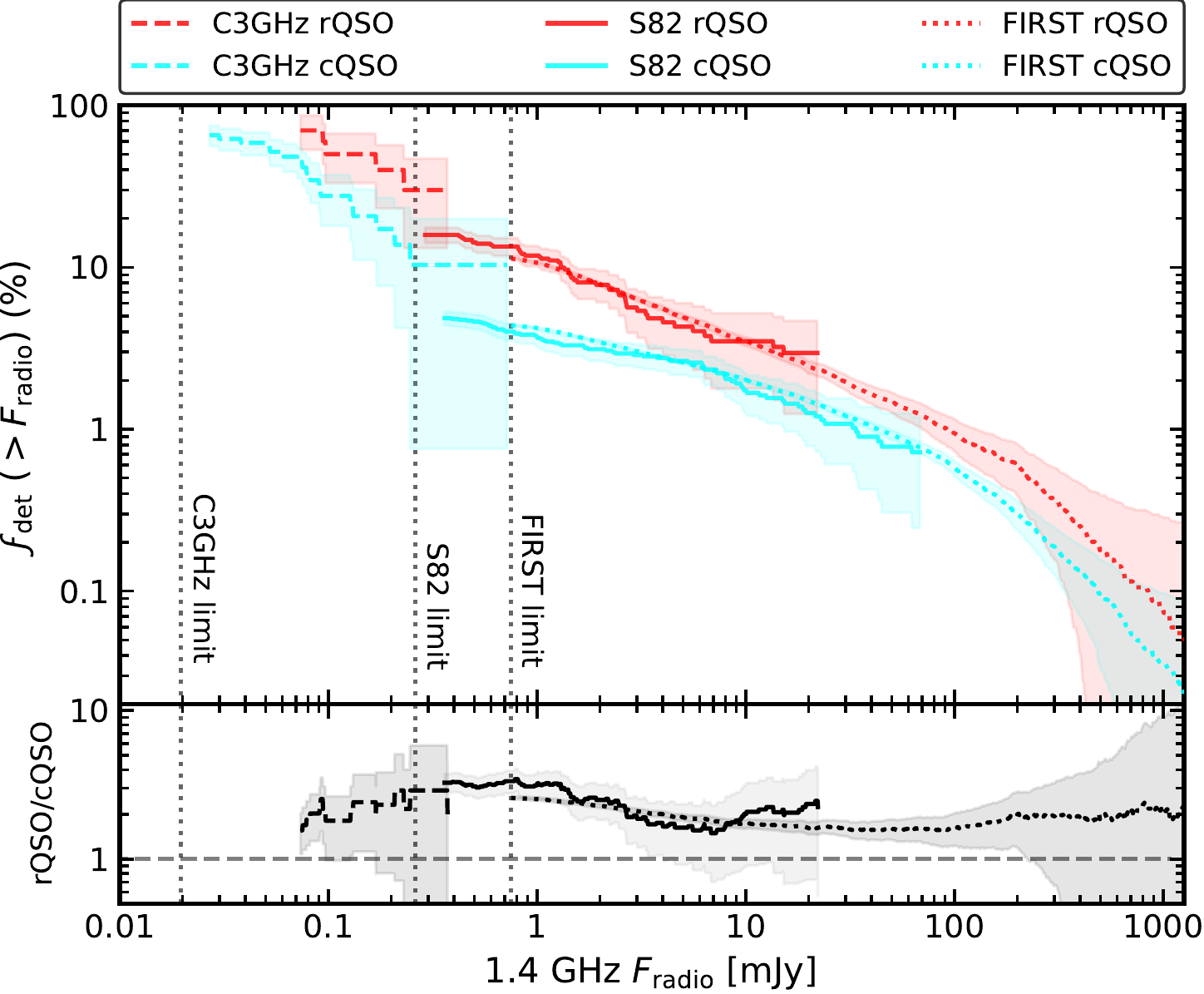}
    \caption{Cumulative radio-detection fractions for the S82, C3GHz and FIRST detected rQSOs (red) and cQSOs (cyan) as a function of radio flux density. The 1.4~GHz flux for the C3GHz sample was calculated from the 3~GHz flux assuming a uniform spectral slope of $\alpha$\,$=$\,$-$0.7 (unless 1.4~GHz data was available). The shaded error region was calculated using the method described in \protect\cite{cam} and corresponds to 1$\sigma$ binomial uncertainties. Cuts were applied at the 5$\sigma$ sensitivity limit of each survey, and data from the bright end was cut when the number of sources became less than 2 (C3GHz) or 10 (S82 \& FIRST) to reduce the widths of the shaded regions; the dotted lines display the 5$\sigma$ flux limits for the three surveys. The bottom panel displays the fractional difference between the radio-detected rQSOs and the radio-detected cQSOs. Across the full range of radio fluxes, the rQSOs have a higher detection fraction than the cQSOs, although the uncertainties are large at the faintest fluxes in the C3GHz survey.}
    \label{fig:det}
\end{figure*}

To explore the radio-detection enhancement in more detail we plotted the cumulative detection fractions of the rQSOs and cQSOs, utilising the C3GHz, S82 and FIRST data as a function of radio flux; see Fig.~\ref{fig:det}. We find that rQSOs exhibit a higher radio-detection fraction compared to the cQSOs in all three surveys, although in regions with low source statistics this enhancement is consistent with unity within the uncertainties. Down to the S82 flux limit, this enhancement is broadly constant at a factor of $\approx$\,2--3. These results are therefore in quantitative agreement with \cite{klindt} \linebreak(see Figure 4 therein), who found a factor $\approx$\,3 enhancement in the FIRST radio-detection fraction of rQSOs compared to their cQSO and blue QSO samples, but with much higher source statistics at the brighter end. We note the apparent detection fraction discontinuity in the overlapping region of fluxes between C3GHz and S82 is consistent within the large statistical uncertainties (only 3 cQSOs and 3 rQSOs are enclosed by this overlapping region). However there may be small systematic contributions due to different deconvolution and source extraction methods between the two radio surveys which are especially prominent at the faint end of the S82 sample.

\begin{figure}
    \centering
    \includegraphics[width=3.3in]{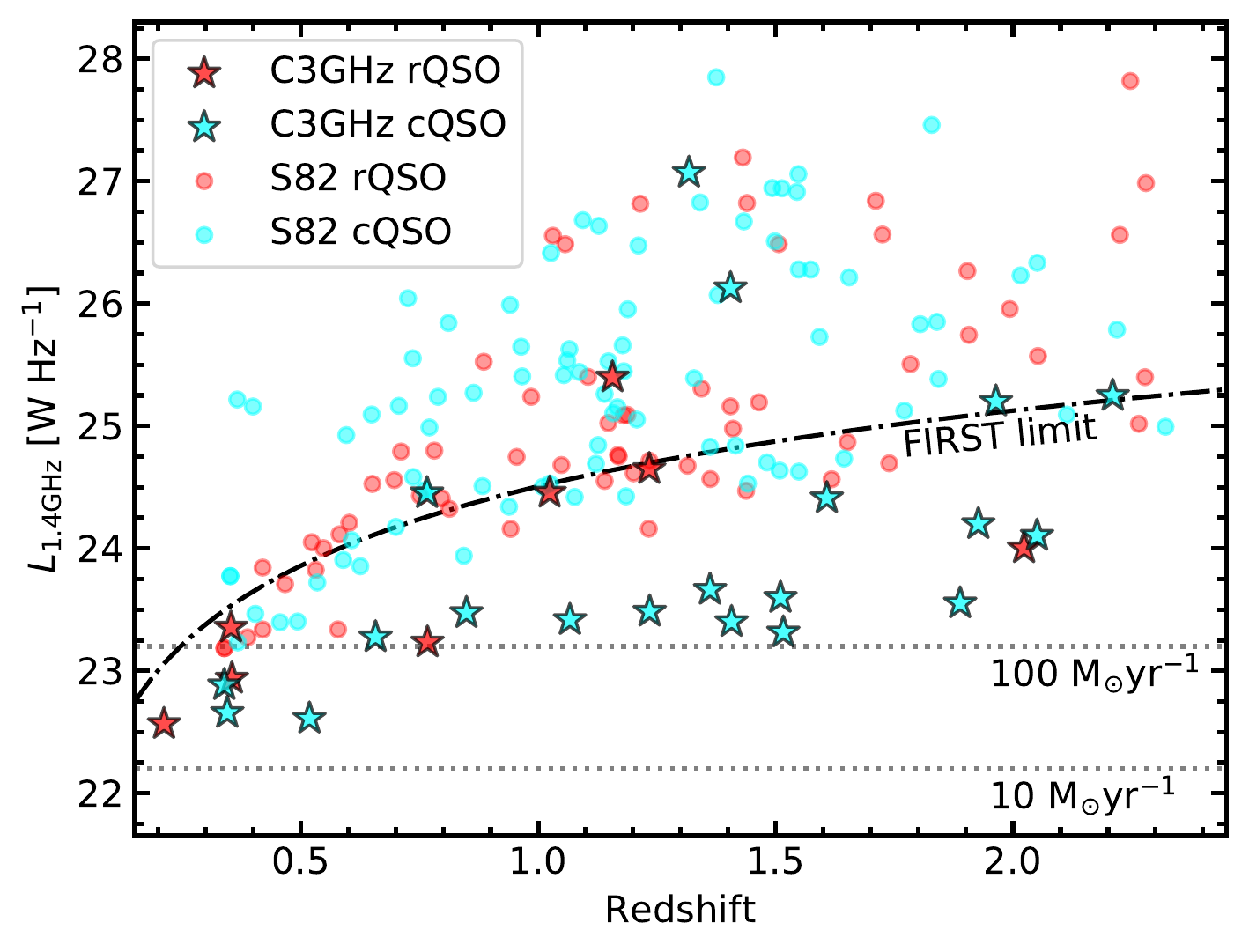}
    \caption{Radio luminosity at 1.4~GHz versus redshift for the S82 (filled circles) and C3GHz (stars) QSOs. The 1.4~GHz luminosity for the C3GHz sample was either taken from the VLA-COSMOS~1.4~GHz catalogue \protect\citep{smol07}, or calculated from the 3~GHz flux by assuming a uniform spectral slope of $\alpha$\,$=$\,$-$0.7. The dotted lines indicate the radio luminosity representing star-formation rates of 10~M$_{\odot}$yr$^{-1}$ and 100~M$_{\odot}$yr$^{-1}$, converted using the \protect\cite{ken} relation. The dot-dashed curve displays the FIRST $5\sigma$ detection threshold; the majority of the C3GHz sources and $\sim$\,25\% (36/143) of the S82 sources lie below this curve.}
    \label{fig:redshiftlum}
\end{figure}

Fig.~\ref{fig:redshiftlum} shows the 1.4~GHz radio luminosity (see Section~\ref{sec:radio}) versus redshift for the cQSOs and rQSOs. The FIRST flux threshold is shown for reference, illustrating that many sources from both \linebreak surveys are too faint to be detected by FIRST. To indicate the radio emission that we might expect from powerful star-formation (SF), we plot the 100~M$_{\odot}$yr$^{-1}$ star-formation rate (SFR) line, with mainly low redshift C3GHz sources falling below this line. This analysis suggests that the radio emission from the majority of the radio-detected QSOs is dominated by the AGN, particularly for the S82 QSOs; see Section~\ref{sec:red_emis} for more detailed constraints on the origin of the radio emission.

\begin{figure*}
    \centering
    \includegraphics[width=6in]{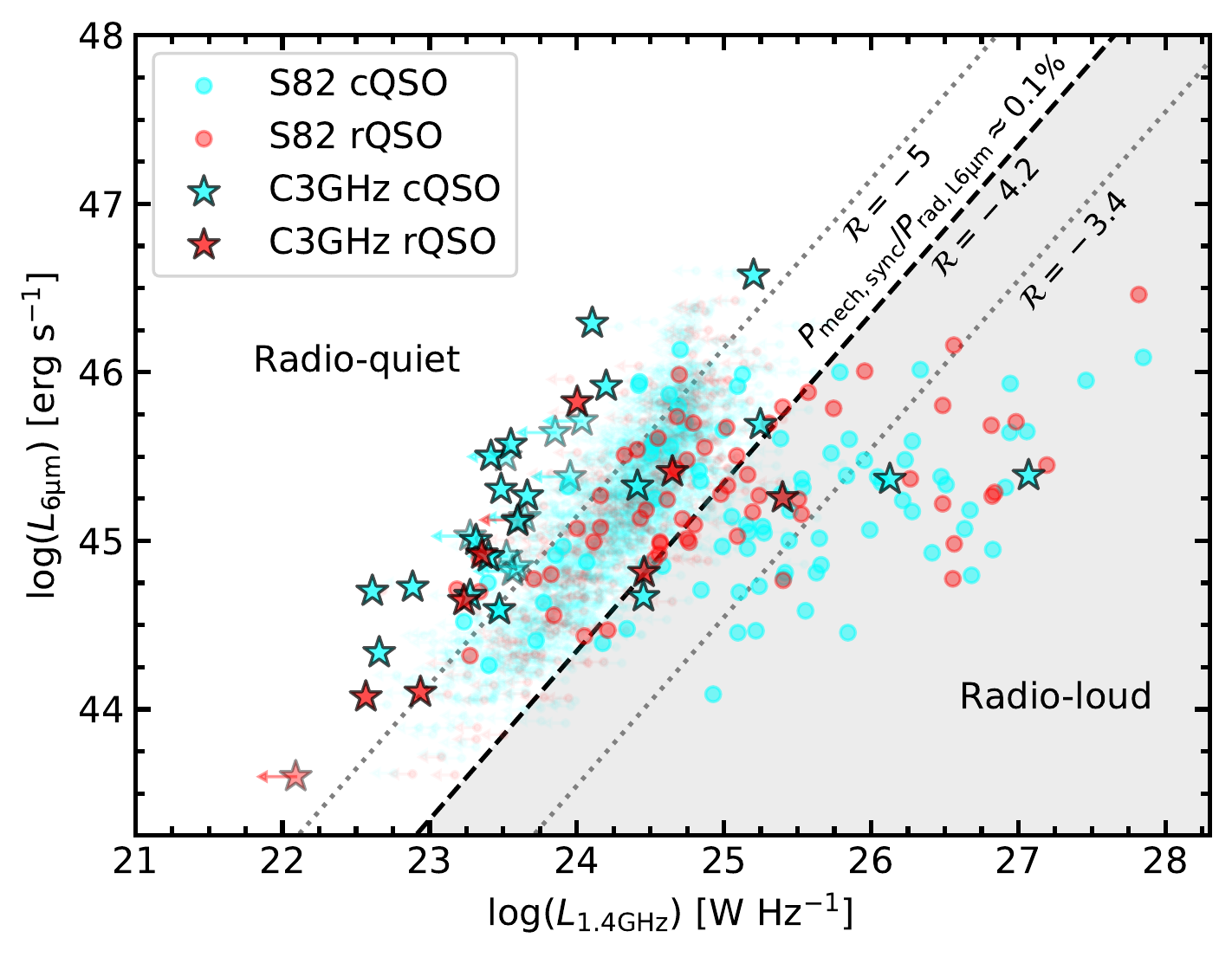}
    \caption{\textit{L}\textsubscript{6$\upmu$m} versus \textit{L}\textsubscript{1.4GHz} for the S82  (filled circles) and C3GHz (stars) sources. The faded points with arrows indicate the properties of the radio-undetected sources, with upper limits derived from the RMS mosaics. The dashed bold line indicates the radio-loud/radio-quiet threshold, defined as a mechanical-to-radiative power of 0.1\% ($\mathcal{R}$\,$=$\,$-$4.2; see Section~3.3 in \citealt{klindt} for details). Additional dotted lines at $\mathcal{R}$\,$=$\,$-$5 and $\mathcal{R}$\,$=$\,$-$3.4 represent the boundaries of the bins used in our analysis, between which we find the highest enhancement, as demonstrated in Fig.~\ref{fig:radio_quiet_sf}.}
    \label{fig:radio_loud}
\end{figure*}

In \cite{klindt}, the enhancement in the radio-detection fraction arose from systems around the radio-loud/radio-quiet threshold. To quantify how many of the rQSOs and cQSOs in our sample are `radio-quiet' we adopted the same ``radio-loudness'' parameter ($\mathcal{R}$) as that used in \cite{klindt}, defined as the dimensionless quantity:
\begin{equation}
\mathcal{R}=\textrm{log\textsubscript{10}}\frac{1.4\times10^{16}\textit{L}\textsubscript{1.4GHz}[\textrm{W Hz}^{-1}]}{\textit{L}\textsubscript{6$\upmu$m}[\textrm{erg s}^{-1}]} .\
\end{equation}
We also used the same radio-loud/radio-quiet threshold of $\mathcal{R}$\,$=$\,$-4.2$, equivalent to a mechanical-to-radiative power ratio of \textit{P}\textsubscript{mech,sync}/\textit{P}\textsubscript{rad,L6$\upmu$m}$\,\approx$\,0.001, which is broadly consistent with the classical threshold often defined using a 5\,GHz-to-2500\,\AA~flux ratio, but is less susceptible to obscuration from dust (see \citealt{klindt} for full details). In Fig.~\ref{fig:radio_loud} we plot \textit{L}\textsubscript{6$\upmu$m} versus \textit{L}\textsubscript{1.4GHz} for the S82 and C3GHz parent and colour-selected samples and indicate our adopted radio-loud/radio-quiet threshold. From this plot it is clear that the extreme radio-loud population is dominated by the cQSOs (67\%), which could indicate a transition between the red and control QSO populations. 

\begin{figure}
    \centering
    \includegraphics[width=3.3in]{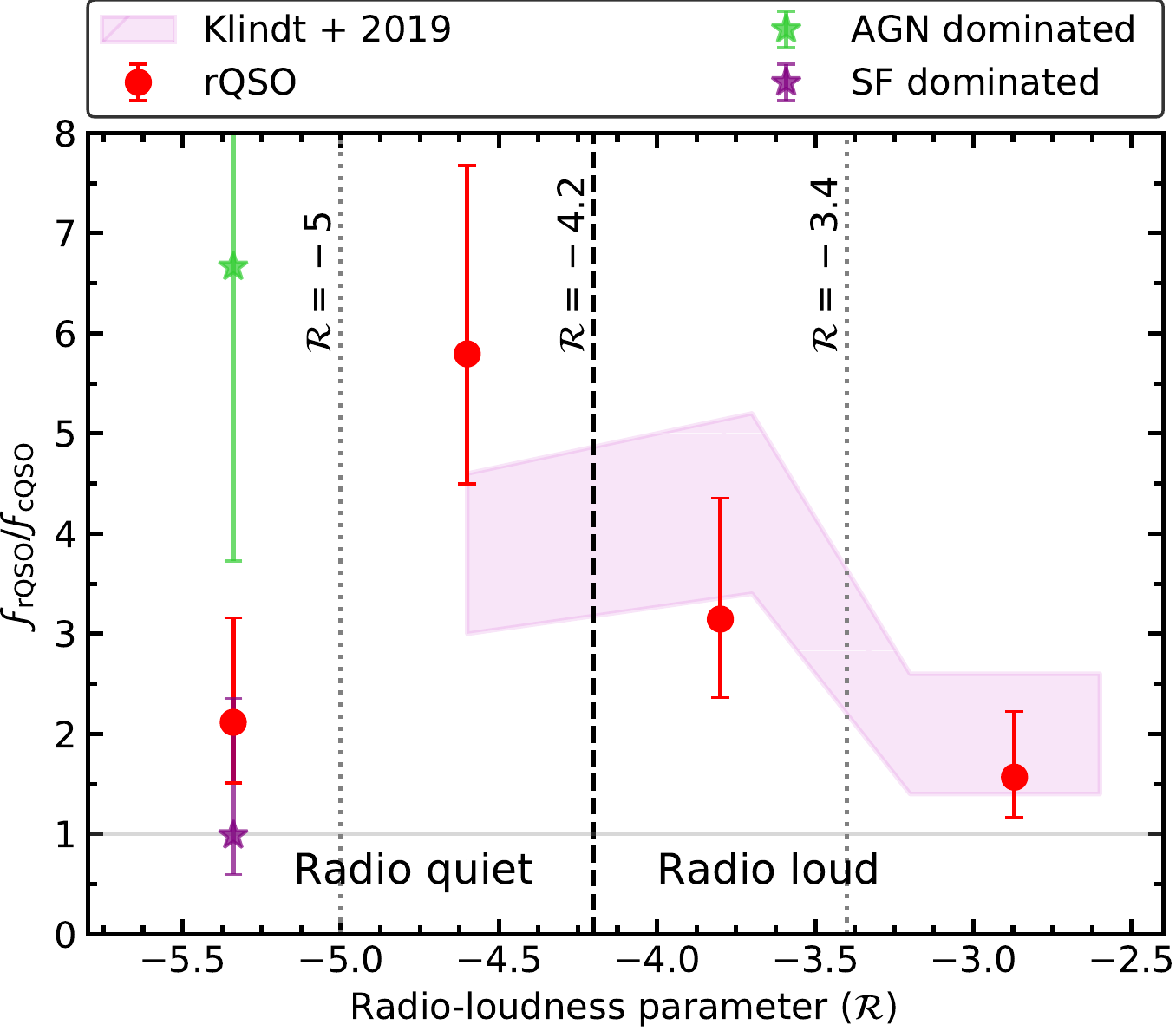}
    \caption{The radio-detection enhancement for rQSOs as a function of the radio-loudness parameter, $\mathcal{R}$ (see Section~\ref{sec:lum}). The red circles give the overall enhancement, which peaks around the radio-loud/radio-quiet threshold, and decreases in both the highest and lowest bin, a trend also seen using deep LOFAR data in our companion study by \protect\cite{rosario}. The shaded pink region represents the result from \protect\cite{klindt}, which is consistent with our data for the overlapping regions of $\mathcal{R}$, within the 1$\sigma$ uncertainties. Therefore we confirm the result at the brighter end, but also push much deeper with the C3GHz data. In the lowest bin ($\mathcal{R}$\,$<$\,$-$5), the C3GHz sources that are SF dominated are indicated by the purple star, and the sources predominately AGN dominated are indicated by the green star (see Section~\ref{sec:red_emis} for details). The vast majority of the SF-dominated sources lie at $\mathcal{R}$\,$<$\,$-$5 suggesting that the radio-detection enhancement seen at $\mathcal{R}$\,$\approx$\,$-$5 to $-$3.4 is due to AGN processes. We argue in Section~\ref{sec:red_emis} that the decrease in the radio enhancement seen at $\mathcal{R}$\,$<$\,$-$5 is therefore not intrinsic but is due to SF diluting the overall radio emission.}
    \label{fig:radio_quiet_sf}
\end{figure}

Splitting the S82 and C3GHz radio-detected sources into four contiguous $\mathcal{R}$ bins (with boundaries $\mathcal{R}$\,$<$\,$-$5, $-$5\,$<$\,$\mathcal{R}$\,$<$\,$-$4.2, $-$4.2\,$<$\,$\mathcal{R}$\,$<$\,$-$3.4 and $\mathcal{R}$\,$>$\,$-$3.4), we calculated the enhancement in the radio-detection fraction of the full colour-selected sample for the rQSOs. From the lowest bin ($\mathcal{R}$\,$<$\,$-$5) to the highest bin ($\mathcal{R}$\,$>$\,$-$3.4), we found an enhancement of 2.1$^{+1.0}_{-0.6}$, 5.8$^{+1.9}_{-1.3}$, 3.1$^{+1.2}_{-0.8}$ and 1.6$^{+0.7}_{-0.4}$, respectively which is displayed in Fig.~\ref{fig:radio_quiet_sf}. The shaded pink region compares our results to \cite{klindt}, who probe down to $\mathcal{R}$\,$\sim$\,$-$4.5. We find good agreement with the enhancement seen in \cite{klindt} (within 1$\sigma$ uncertainties), despite using an optically fainter QSO sample, but also show that with C3GHz we can push to the levels where SF is likely to be important (detection limit of C3GHz at $z$\,$=$\,1.5: 3.9\,$\times$\,$10^{23}$~WHz$^{-1}$, see Fig.~\ref{fig:redshiftlum}). Therefore we confirm that the radio enhancement seen in rQSOs appears to arise around the radio-loud/radio-quiet threshold with tentative evidence for a decrease at both $\mathcal{R}$\,$<$\,$-$5 and $\mathcal{R}$\,$>$\,$-$3.4; the latter confirmed with better source statistics in \cite{klindt}. This is also confirmed with a larger sample using deep LOFAR data, which gives qualitatively the same result, demonstrating there is a decrease in the radio enhancement of rQSOs for $\mathcal{R}$\,$<$\,$-$5 \citep{rosario}.

\subsection{Median radio stacking}\label{sec:stack}
The C3GHz data probes radio flux values for individually detected sources down to $\upmu$Jy levels, allowing for the majority of the QSOs to be detected in the radio band. However, C3GHz only covers a small area with modest source statistics, whereas S82 covers a $\sim$\,35 times larger area, resulting in a higher number of sources, albeit 12 times shallower. Radio stacking allows us to compare the average properties of undetected S82 QSOs that would have been detected at the C3GHz depth.

\begin{figure}
    \begin{center}
    \includegraphics[width=3.2in]{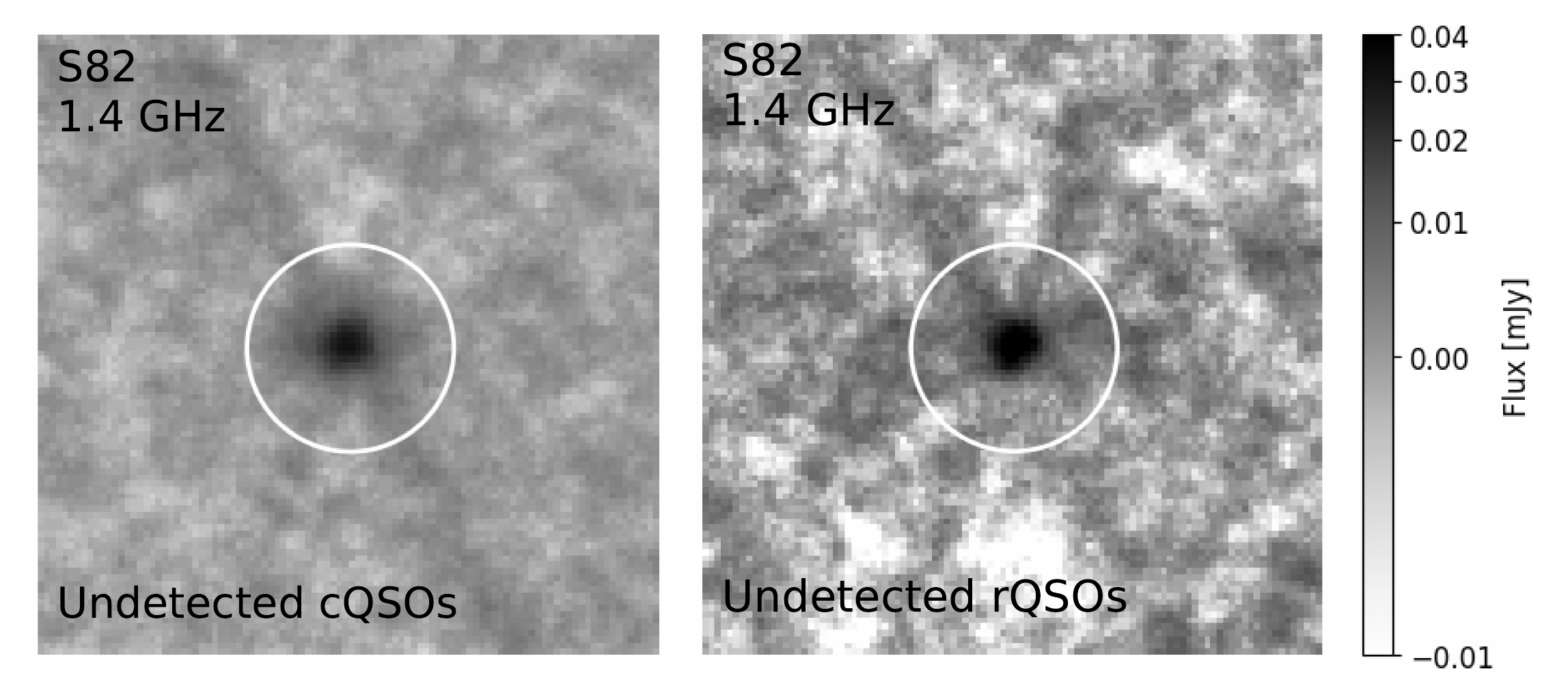}
    \end{center}
    \caption{The resulting $0\farcm5\times 0\farcm5$ images of the median radio (1.4~GHz) stacked radio-undetected cQSOs (left) and radio-undetected rQSOs (right) in the S82 field. The white circle has a 10$''$ radius and the colour bar has an \textit{arcsinh} scaling. A two-dimensional Gaussian was fitted to the images giving a peak 1.4~GHz flux density of 35.0\,$\pm$\,2.7~$\upmu$Jy and 26.1\,$\pm$\,1.8~$\upmu$Jy for the S82 rQSOs and cQSOs, respectively, although this has not been corrected for any CLEAN or snapshot bias (see \protect\citealt{white06}).}
    \label{fig:stack}
\end{figure}

We stacked the S82 data of the radio-undetected QSOs following the procedure defined in Section~\ref{sec:median}. We found median 1.4~GHz flux values of 35.0\,$\pm$\,2.7~$\upmu$Jy (S/N~$\sim$\,13) and 26.1\,$\pm$\,1.8~$\upmu$Jy (S/N~$\sim$\,15) for the S82 rQSOs and cQSOs, respectively. Again, fluxes on the computed stacks are not corrected for any CLEAN or snapshot bias, but we can make a qualitative comparison between the stacks. The final stacked images for the S82 radio-undetected colour-selected QSOs are shown in Fig.~\ref{fig:stack}, displaying the significant detections. Although the difference is small, the rQSOs appear to be relatively brighter in the radio band than the cQSOs for the undetected population. It is important to note that these radio flux values may be a combination of AGN activity and SF, especially for the C3GHz stack (see Section~\ref{sec:red_emis}).

We also analysed the redshift dependence of our radio-undetected stacks. Splitting the S82 radio-undetected cQSOs and rQSOs into four contiguous redshift bins with boundaries 0.2, 0.5, 0.8, 1.5 and 2.4, we calculated the 1.4~GHz flux at the median redshift of all QSOs in each bin. For all redshift bins, the undetected rQSOs have a slightly higher relative flux compared to the cQSOs.
\subsection{Radio morphology fractions of red and control QSOs}\label{sec:morph}
Using the FIRST radio data, \cite{klindt} found that the radio-detection enhancement of their rQSOs was due to a higher incidence of compact and faint radio counterparts. On the basis of the FIRST data, sources were classified as compact if their radio emission is unresolved on scales less than the 5$''$ beam size (i.e.\ corresponding to 43~kpc at $z$\,$=$\,1.5). Using the higher spatial resolution radio data in this study, we can probe down to host-galaxy scales (S82: 16~kpc at $z$\,$=$\,1.5; C3GHz: 7~kpc at $z$\,$=$\,1.5), resolving finer-scale structure in many of the FIRST compact sources and testing whether rQSOs still show an enhancement in compactness on these smaller scales. The greater depth of our radio data also allows us to determine the morphology of many systems too faint to detect or categorise with FIRST.

\begin{table}
    \caption{The number of S82 rQSOs and cQSOs in each of the morphology classes, with the overall fraction of the colour-selected sample shown. The fractional difference between the rQSOs and the cQSOs is 4.2$^{+1.3}_{-0.9}$ for the faint category, 2.9$^{+0.9}_{-0.6}$ for the compact category and 2.2$^{+1.6}_{-0.8}$ for the extended category.}
    \begin{center}
    \begin{tabular}{p{2cm}| p{1cm} p{1cm}| p{1cm} p{1cm}}        \hline
        \hline
        \multirow{2}{*}{Classification}&\multicolumn{2}{c|}{rQSO} & \multicolumn{2}{c}{cQSO} \\
        &  Num. & \% & Num. & \% \\
        \hline
        Faint & 28 & 7.5$^{+1.6}_{-1.2}$\ & 30 & 1.8$^{+0.4}_{-0.3}$  \\ 
        Compact & 24 & 6.5$^{+1.5}_{-1.1}$ & 34 & 2.0$^{+0.4}_{-0.3}$ \\
        Extended & 6 & 1.6$^{+0.9}_{-0.4}$ & 12 & 0.7$^{+0.3}_{-0.2}$\\
         FRII-like & 3 & 0.8$^{+0.8}_{-0.2}$ & 6 & 0.4$^{+0.2}_{-0.1}$ \\
        \hline
        \hline
    \end{tabular}
    \end{center}
    \label{tab:morph_num} 
\end{table}

\begin{figure*}
    \centering
    \includegraphics[width=6in]{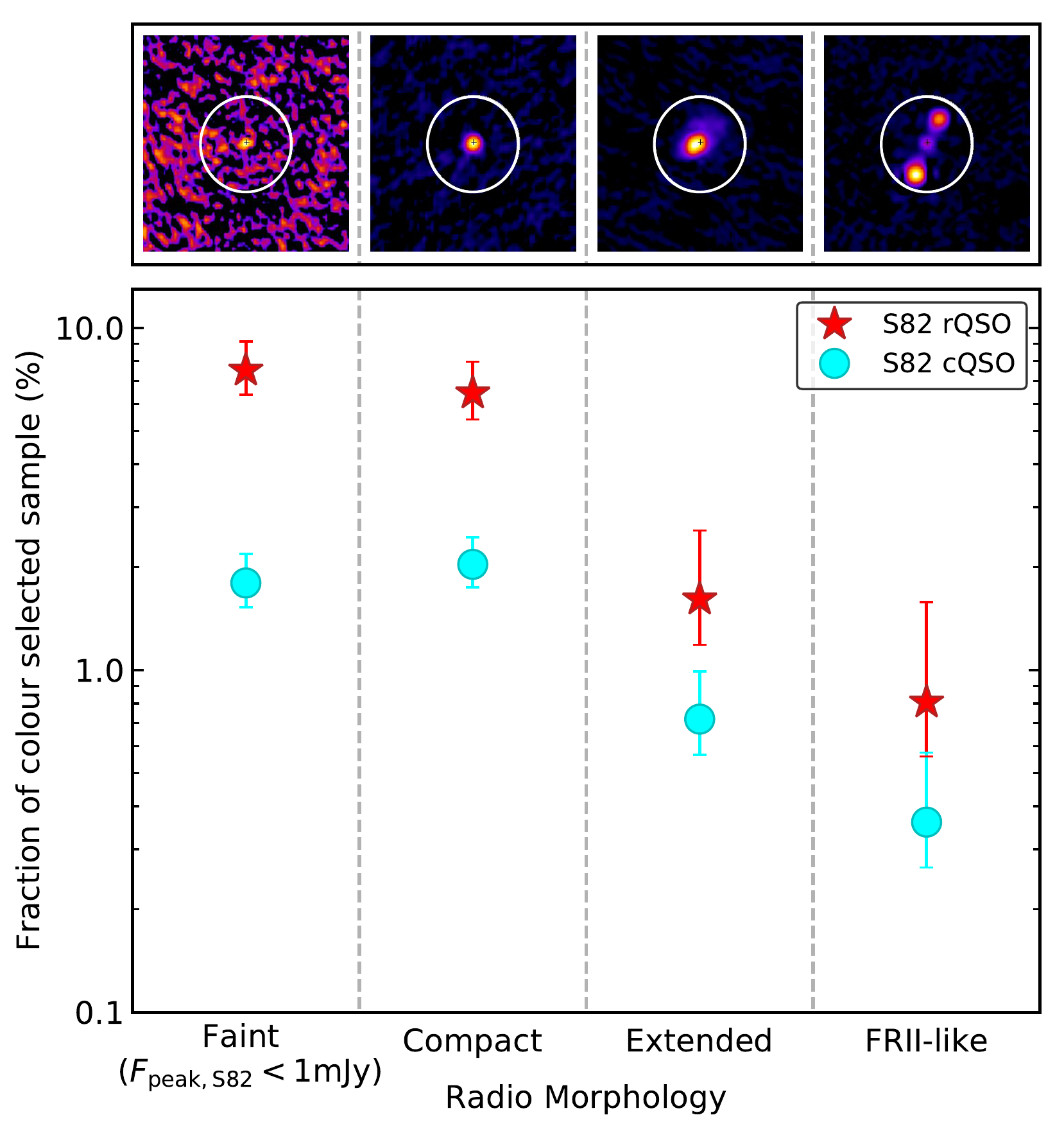}
    \caption{Radio morphology fractions for the S82 rQSOs and cQSOs. The thumbnails display 45$''$\,$\times$\,45$''$ examples of the four morphology types for the S82 VLA data \protect\citep{hodge}. The white circles represent a 10$''$ radius. The error bars were calculated using the method described in \protect\cite{cam} and correspond to the equivalent 1$\sigma$ uncertainties on the ratio. The rQSOs show a strong factor $\approx$\,3 enhancement in the faint and compact classes and a less significant factor $\approx$\,2 enhancement in extended radio morphologies; however, there are no differences between the rQSOs and cQSOs in the FRII-like group within the 1$\sigma$ error bars although we note that there are only 9 FRII-like QSOs in total; however, we note that we also did not see any significant difference in the FRII-like category between rQSOs and cQSOs in our companion study \protect\citep{klindt} which had significantly improved source statistics.}
    \label{fig:morph}
\end{figure*}

We classified the radio morphologies of the rQSOs and cQSOs following Section~\ref{sec:morph_meth}, using the high-resolution S82 and C3GHz radio data. The number of objects classified in our four morphology categories is shown in Table~\ref{tab:morph_num}, with the fractions calculated from the colour-selected QSOs within Stripe 82. Fig.~\ref{fig:morph} shows the S82 radio morphology fractions: the rQSOs show a clear enhancement in the faint and compact morphology groups by a factor of 4.2$^{+1.3}_{-0.9}$ and 2.9$^{+0.9}_{-0.6}$, respectively, when compared to the cQSOs, \linebreak quantitatively consistent with the result from \cite{klindt}. However, we now also see a tentative difference in the extended group, with rQSOs showing a factor 2.2$^{+1.6}_{-0.8}$ enhancement compared to the cQSOs. This enhancement is not seen at the lower resolution of FIRST, where \cite{klindt} found the two samples to be consistent in the extended category (defined as radio emission on scales larger than 5$''$). Although this enhancement is around the 2$\sigma$ significance level, this may indicate fundamental differences between rQSOs and cQSOs on 1$\farcs$8--5$''$ scales (16--40~kpc at $z$\,$=$\,1.5) that can be explored with even higher spatial resolution. There is no difference in the FRII-like category within the 1$\sigma$ error bars, consistent with that found in \cite{klindt}.

As shown in Fig.~\ref{fig:radio_quiet_sf}, we found the biggest enhancement in the radio-detection fraction for rQSOs arises around the radio-loud/radio-quiet threshold ($-$5\,$<$\,$\mathcal{R}$\,$<$\,$-$3.4). Looking at the morphologies of sources in these bins, we find that the radio enhancement is entirely driven by compact and faint sources, in agreement with \cite{klindt} but pushing to even lower $\mathcal{R}$ values.

\begin{figure*}
    \centering
    \includegraphics[width=6in]{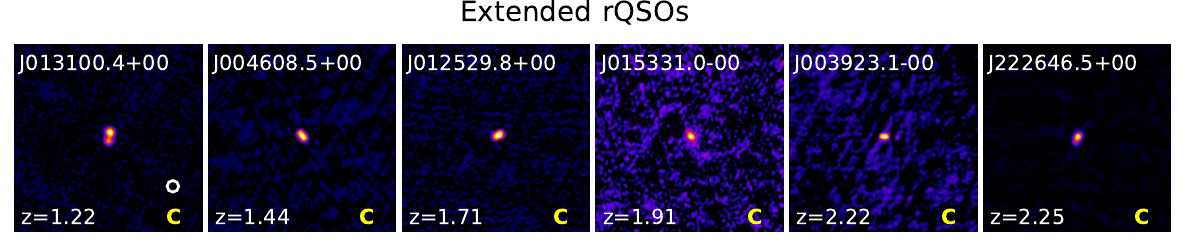} \\
    \vspace{5mm}
    \includegraphics[width=6in]{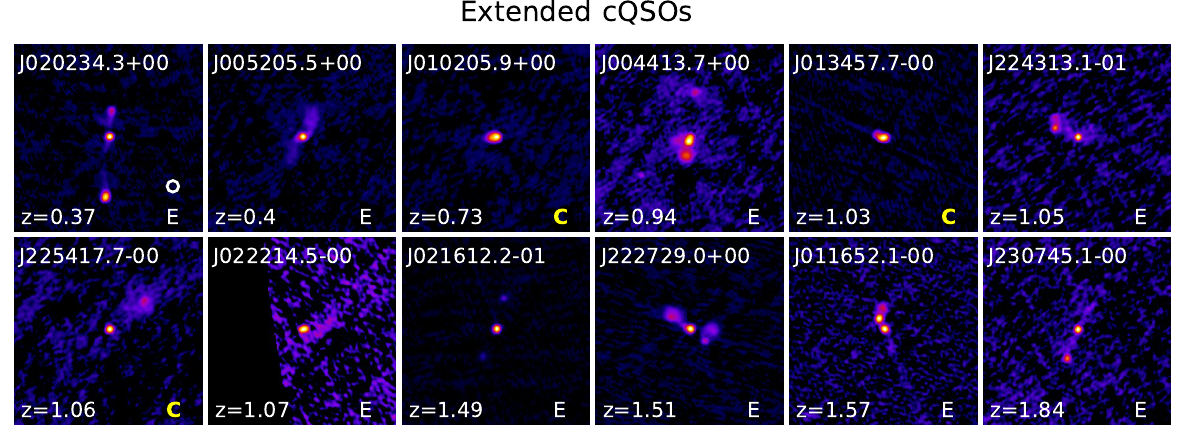}
    \caption{Thumbnails (1$'$\,$\times$\,1$'$) of S82 extended rQSOs (top) and cQSOs (bottom) using the S82 data \protect\citep{hodge}. The white circle in the first cutout represents the 1$\farcs$8 beam size. The redshift is shown in the bottom left corner and the VLA name is shown in the top left corner. The classification from visually inspecting the FIRST data \protect\citep{first} is shown in the bottom right corner of the image; C (compact) or E (extended). The sources that were reclassified from FIRST with the higher resolution data are highlighted in yellow, showing that 0/6 of the rQSOs were classified as extended in the FIRST data, compared to 9/12 of the cQSOs which indicates that the rQSOs are only extended on $\approx$\,1$\farcs$8--5$''$ scales.}
    \label{fig:ext}
\end{figure*}

Fig.~\ref{fig:ext} shows cutouts of the S82 QSOs with extended radio morphologies, all of which are radio-loud and detected by FIRST. The FIRST classification is shown in the bottom right corner of each cutout, indicating which sources are reclassified in the S82 data. Many of the S82 extended sources were FIRST compact sources where we can now resolve more diffuse emission. On the basis of the FIRST morphology classifications, none of the rQSOs classified as extended in the S82 data were found to show extension beyond the 5$''$ resolution of the FIRST data. This is compared to 75\% (9/12) of the cQSOs, indicating that the radio emission of the extended rQSOs is on smaller scales than the cQSOs ($\approx$\,1$\farcs$8--5$''$; 16--43~kpc); indeed, although limited by small source statistics, our data suggests a factor $\approx$\,10 times more rQSOs are extended on these scales than cQSOs. 

We also investigated the radio morphology fractions using the C3GHz data: the rQSOs still show an enhancement in the compact category of 2.5$^{+1.0}_{-0.9}$ compared to the cQSOs (7/8 compact rQSOs and 8/20 compact cQSOs), but now going down to radio fluxes $\approx$\,12 times fainter and a spatial resolution $\approx$\,2.4 times higher than that of S82. There are no significant differences in the faint and extended groups, however there are too few sources to draw any statistically significant conclusions. 
\section{Discussion}\label{sec:dis}
We have used high-resolution, deep radio data to explore differences between red QSOs and typical QSOs detected in the SDSS DR14 quasar survey. Using the VLA Stripe~82 and VLA-COSMOS 3~GHz data, we have explored the radio properties of QSOs to a depth of $\sim$\,3 and 30 times that of FIRST, down to scales of 16~kpc and 7~kpc at $z$\,$=$\,1.5, respectively. With these data we confirm the results from \cite{klindt}, but down to lower radio luminosities and hence unexplored regions of the radio-loudness plane (\textit{L}\textsubscript{6$\upmu$m} vs \textit{L}\textsubscript{1.4GHz}). From this we can gain new insight into the origin and scale of the radio emission in radio-quiet systems.

We find a significant enhancement in the radio emission of red QSOs down to low radio fluxes ($\sim$\,0.3~mJy in S82). Looking at the radio-detection fraction as a function of flux for FIRST, S82 and C3GHz, we find an enhancement that is broadly constant at a factor of $\approx$\,3, although the uncertainties are large at faint radio fluxes in the C3GHz field (see Fig.~\ref{fig:det}). We pushed this limit further by stacking the radio-undetected S82 rQSO and cQSO sources, which resulted in a higher average relative flux for the rQSOs. We also explored the radio-detection enhancement in the rQSOs as a function of the radio-loudness parameter $\mathcal{R}$, where we confirm the enhancement peaks around the radio-quiet threshold $-$5\,$<$\,$\mathcal{R}$\,$<$\,$-$3.4 (enhancement of $\sim$\,3--6); however, for our lowest bin ($\mathcal{R}$\,$<$\,$-$5), which probes a magnitude below that explored in \cite{klindt}, we see a decrease in the enhancement by a factor of $\sim$\,3. This suggests that the fundamental differences between rQSOs and cQSOs become apparent at $\mathcal{R}$\,$<$\,$-$3.4, possibly due to a different process starting to dominate the radio emission. However, whether this process is due to AGN-driven winds, frustrated jets, star-formation or a coronal component (to name a few) is unclear \citep{zak_gren,kell,laor,pan,jarvis}. 

An enhanced fraction of broad absorption line QSOs (BALQSOs; known to host powerful winds) have been found in red QSOs \citep{Urrutia_2009}, which could be evidence that red QSOs are more wind dominated than typical QSOs. Although BALQSOs are virtually all radio-quiet, they have also been found to show a remarkably similar enhancement in the radio as that found for red QSOs \citep{klindt,balqso}. Further corroborating but independent evidence for wind-dominated sources becoming more relevant at lower $\mathcal{R}$ values comes from \cite{med}, who showed a significant inverse correlation between the X-ray measured column density of the ionised wind in AGN and the radio-loudness parameter. On the basis of this we may therefore expect a larger fraction of the rQSOs to have more extreme winds than typical QSOs, which would be consistent with rQSOs representing a younger phase in the QSO evolutionary scenario. 

To try and constrain the nature of the differences in the radio emission we investigated the radio morphologies. Probing down to arcsecond scales and sensitivity limits at least a factor $\approx$\,3 deeper than FIRST has allowed us to undertake a thorough test of the morphology result from \cite{klindt}, who found that rQSOs showed an enhancement in the compact and faint radio morphology classes when compared to typical QSOs; we find that $\sim$\,40\% (24/61) of our rQSOs are compact on scales of $\sim$\,16~kpc at $z$\,$=$\,1.5, the resolution of our data. We found good agreement with their result, but we also found a tentative enhancement (at the 2$\sigma$ statistical level) in the extended morphologies of rQSOs which is not seen at the resolution of FIRST, suggesting that we are starting to resolve the scales of the radio structures that are driving these differences (16--43~kpc at $z$\,$=$\,1.5).

We found that none of the extended rQSOs are classified as extended at the resolution of FIRST, which again indicates that the radio emission in rQSOs occurs on scales 16--43~kpc (at $z$\,$=$\,1.5). This enhancement in the extended rQSOs is driven by the improved resolution of the survey, with sources that may have been classified as compact at the resolution of FIRST now reclassified as extended (see Appendix~\ref{sec:first}). The compactness of the radio emission could imply that these objects are in a young phase compared to typical bluer QSOs, which are more likely to have extended radio jets, consistent with an evolutionary scenario. Using data from \cite{Siem}, who investigate young radio sources with known kinematic radio jet ages, a crude linear relationship between jet age and the size of the radio emission can be constructed. Applying this relation to the scales on which these radio differences occur (16--43~kpc at $z$\,$=$\,1.5) yields a range of 0.2--0.5~Myrs as a rough estimate for the age of the extended radio emission in the rQSOs. We note that no correction for orientation has been made, and so this represents a lower limit; however, this timescale is broadly consistent with other work that estimate a rQSO phase duration of a few Myrs \citep{hop6,glik12}.

Overall our results add weight to the emerging picture that red QSOs are fundamentally different to typical QSOs and cannot be explained by the simple orientation model alone. These differences in the radio properties are most likely to be driven by differences in the accretion or ``environment'' (see Footnote 1); since \linebreak \cite{klindt} demonstrated that differences in the accretion appear unlikely, environmental differences are the most probable cause. Below, we focus our discussion on the origin of the radio emission in these systems (Section~\ref{sec:red_emis}) and constrain the SF contribution for the C3GHz sample (Section~\ref{sec:red_emis_overall}). To maximise the source statistics we use the S82 and C3GHz parent samples rather than focus on the colour-selected QSOs (see Table~\ref{tab:rad_num}); however, we do explore whether SF can explain the rQSO enhancement\footnote{We note that our results in Sections~\ref{sec:red_emis} and \ref{sec:red_emis_overall} also hold when restricting the analysis to the colour-selected samples, but with poorer source statistics.}.

\subsection{SF contribution to the radio emission in radio-quiet QSOs}\label{sec:red_emis}
Using the deeper data, we find that the radio enhancement seen in rQSOs is stronger for sources around the radio-loud/radio-quiet threshold ($-$5\,$<$\,$\mathcal{R}$\,$<$\,$-$3.4), peaking in the bin just below the threshold and dropping at lower and higher values of $\mathcal{R}$ (see Section~\ref{sec:lum}). For $\mathcal{R}$\,$>$\,$-$4.2, all C3GHz and S82 parent sample QSOs have \textit{L}\textsubscript{1.4GHz~}$\gtrsim 2\times 10^{24}$~WHz$^{-1}$, which corresponds to an equivalent SFR of $\gtrsim1000$~M$_{\odot}$yr$^{-1}$, suggesting that the vast majority are likely to be AGN dominated. However at $\mathcal{R}$\,$<$\,$-$4.2, it is unclear whether SF or AGN processes are the origin of the radio enhancement. We can make progress in understanding the origin of the radio enhancement by utilising the \textit{Herschel} far-IR (FIR) data in COSMOS and Stripe 82 and constrain the contribution from SF.

\begin{figure}
    \centering
    \includegraphics[width=3.3in]{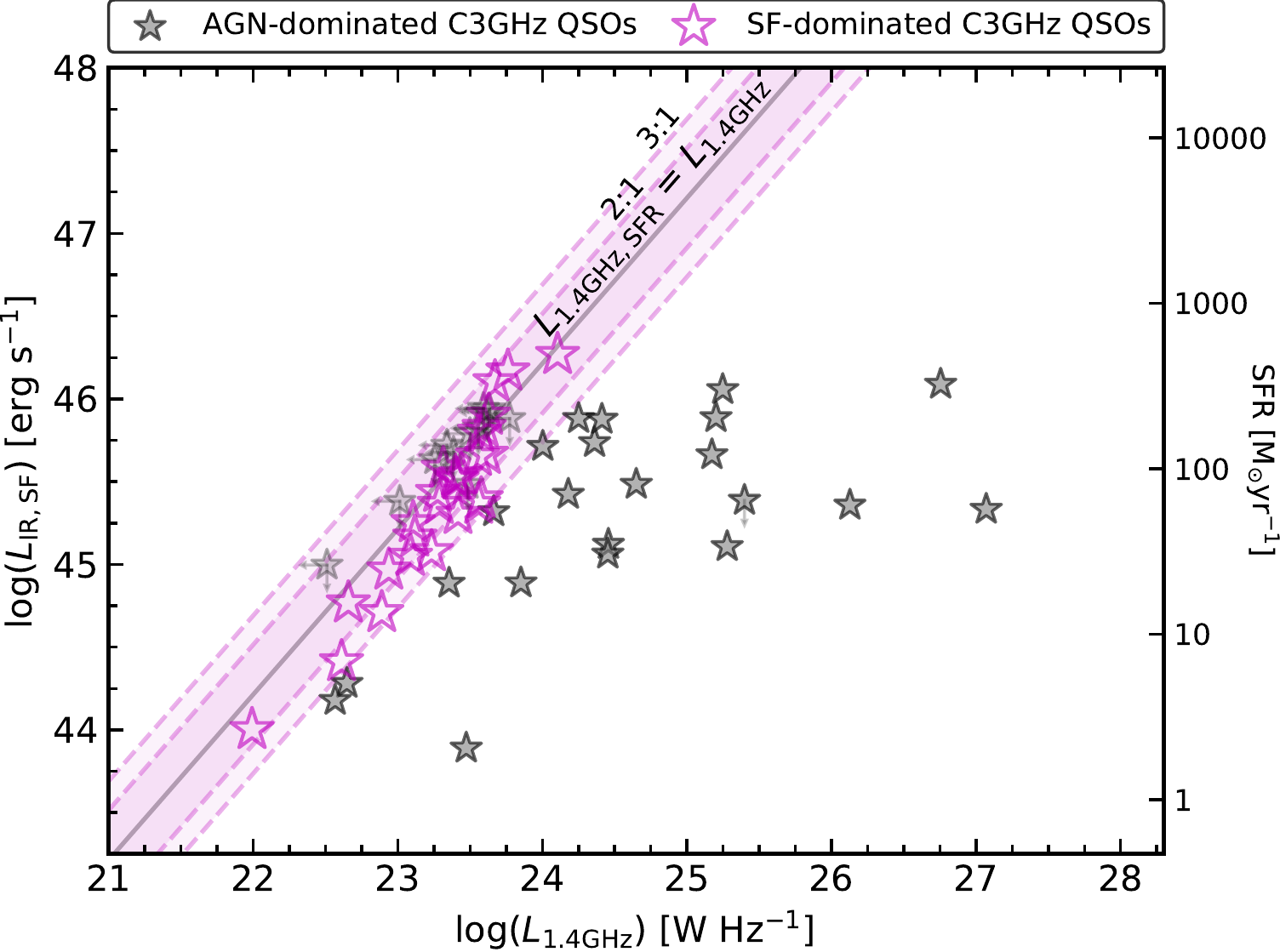}
    \caption{IR luminosity from star-formation versus 1.4~GHz luminosity for the C3GHz parent QSOs. Only QSOs with \textit{Herschel} coverage are shown (61/70 C3GHz parent sample QSOs). The radio luminosity upper limits were calculated from the 3~GHz RMS mosaics, and are converted to 1.4~GHz assuming a uniform spectral index of $-$0.7. The grey solid line indicates a 1:1 relation between the measured 1.4~GHz radio emission and that expected from the SFR of a source. The SFR was calculated from the IR luminosity using the \protect\cite{ken98} conversion and scaling to a Chabrier IMF. The shaded purple region illustrates a factor of three deviation from this line, the range adopted here for a source classified as SF dominated. Excluding sources with an upper limit on the SFR, 27 QSOs lie within this region, illustrated by empty purple stars, giving an estimate of the fraction of QSOs with radio emission that is potentially SF dominated of 44\%.}
    \label{fig:sf}
\end{figure}

In this analysis we utilised rest-frame (8--1000~$\upmu$m) IR luminosity constraints on SF from \cite{stan15} (covering 53/70 of our C3GHz parent QSO sample), and \cite{smol17} (covering 47/70 of our C3GHz parent QSO sample), who both fit an AGN and SF component to the SEDs of X-ray and radio AGN, respectively. The IR luminosities were taken preferentially from \cite{stan15}, resulting in a total of 61/70 of our C3GHz parent sample with IR data from one of the two catalogues. Since S82 is much shallower than C3GHz and does not have deep \textit{Herschel} data, we only use this sample to probe the brighter IR end of the QSO population.

The tight radio-FIR relationship for star-forming galaxies provides a way to identify SF-dominated QSOs. In Fig.~\ref{fig:sf}, we plot the SF luminosity constraints of 61 C3GHz parent QSOs with IR data. We also plot on the line indicating a 1:1 relation between the 1.4~GHz radio emission expected from the SFR of a source, which was calculated by converting the AGN host component of the IR luminosity to a radio luminosity using the relation given in \cite{delhaize}, assuming $q$\textsubscript{TIR}\,$=$\,2.64 \citep{bell}. This relation was calibrated on the VLA-COSMOS~3~GHz data and so is suitable for our sample. We define the origin of the radio emission as dominated by SF if the measured radio luminosity is within a factor of 3 of the 1:1 relation; resulting in 27/61 ($\approx$\,44\%) sources classified as SF dominated. If a source has an upper limit on the SFR then the origin of the radio emission is classified as uncertain. 

\begin{figure*}
    \centering
    \includegraphics[width=6in]{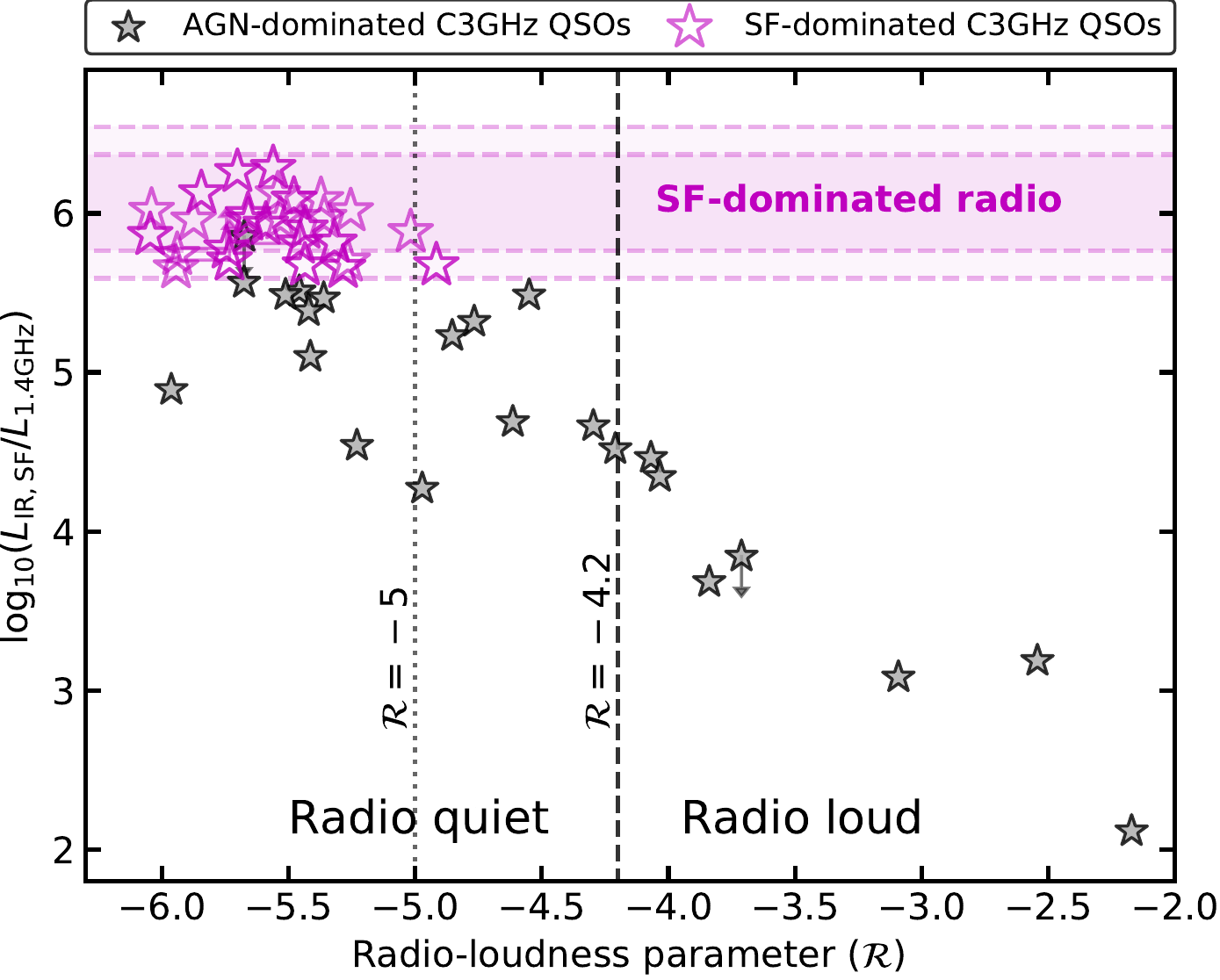}
    \caption{\textit{L}\textsubscript{IR,SF}/\textit{L}\textsubscript{1.4GHz} vs $\mathcal{R}$ for the C3GHz parent sample. Only C3GHz QSOs with \textit{Herschel} coverage and a detection in either the IR or radio are shown (50/70 C3GHz parent sample QSOs). We note that all the QSOs with an upper limit on both \textit{L}\textsubscript{IR,SF} and \textit{L}\textsubscript{1.4GHz} lie at $\mathcal{R}$\,$<$\,$-$5. The grey stars represent AGN-dominated C3GHz QSOs and the empty purple stars represent the SF-dominated QSOs. The arrows indicate upper limits on either \textit{L}\textsubscript{IR,SF} or \textit{L}\textsubscript{1.4GHz}. The purple shaded region displays our selection of a SF-dominated source, as shown in Fig.~\ref{fig:sf}.}
    \label{fig:radio_loud_sf}
\end{figure*}

In order to further explore the origin of the radio emission down to low $\mathcal{R}$ values, in Fig.~\ref{fig:radio_loud_sf} we plot \textit{L}\textsubscript{IR,SF}/\textit{L}\textsubscript{1.4GHz} as a function of $\mathcal{R}$ for the C3GHz parent sample QSOs with IR coverage. Our selection of SF-dominated QSOs (from Fig.~\ref{fig:sf}) is shown as the shaded purple horizontal region and the dashed vertical line indicates the radio-loud/radio-quiet threshold ($\mathcal{R}$\,$=$\,$-$4.2). The QSOs with radio emission dominated by SF are indicated by the empty purple stars. As is clear from this figure, all but one of the SF-dominated QSOs lie at $\mathcal{R}$\,$<$\,$-$5, but it is also apparent that not all QSOs that lie below this line will have radio emission dominated by SF. This suggests that SF only significantly contaminates the radio emission in QSOs at $\mathcal{R}$\,$\lesssim$\,$-$5. Since the C3GHz data only covers a small area, it will miss the most extreme star-forming QSOs: therefore to test whether this boundary in \textit{L}\textsubscript{6$\upmu$m}--\textit{L}\textsubscript{1.4GHz} does indeed isolate the majority of the star-forming population, we take a sample of 148 \textit{Herschel}-observed QSOs in Stripe 82 (sample from \citealt{Dong}), requiring $W1$, $W2$, $W3$ SNR\,$>$\,2 and 0.2\,$<$\,$z$\,$<$\,2.4. We calculated \textit{L}\textsubscript{6$\upmu$m} for these QSOs following the same approach as Section~\ref{sec:six}. After removing the AGN contribution to the FIR luminosity, \cite{Dong} then convert this to a SFR using the \cite{ken98} equation. After converting the SFRs into radio emission using the \cite{ken} relation, we find 94\% (139/148) of the sample have $\mathcal{R}$\,$<$\,$-$5 (full range: $-$6.6\,$<$\,$\mathcal{R}$\,$<$\,$-$4.8). Since we assume that all of the radio emission comes from SF, this gives an upper limit, demonstrating that the radio emission from QSOs with $\mathcal{R}$\,$>$\,$-$5 is dominated by AGN processes.

Using this knowledge, we now try to understand the decrease in the radio-detection fraction for the rQSOs at $\mathcal{R}$\,$<$\,$-$5 (see Section~\ref{sec:lum}); here we split the C3GHz sources at $\mathcal{R}$\,$<$\,$-$5 into those that are SF dominated and those that are AGN dominated. Fig.~\ref{fig:radio_quiet_sf} displays the radio-detection enhancement in the rQSOs for the different bins considered in Section~\ref{sec:lum}. The enhancement in the purely AGN-dominated C3GHz sources is shown as the green star which, although limited by source statistics, is consistent with the enhancement at $-$5\,$<$\,$\mathcal{R}$\,$<$\,$-$3.4. When looking at the purely SF-dominated C3GHz sources (shown as the purple star), the enhancement becomes consistent with unity, suggesting that differences in the radio properties is not predominantly due to SF. Therefore, as AGN processes clearly dominate the radio emission at higher $\mathcal{R}$ values, the decrease in the enhancement seen at $\mathcal{R}$\,$\lesssim$\,$-$5 could be due to an increase in the relative contribution to the radio emission from SF as the AGN component becomes comparably weak; i.e.\ the decrease in the radio enhancement at $\mathcal{R}$\,$\lesssim$\,$-$5 is not intrinsic but is due to SF diluting the overall radio emission. However, larger QSO samples with faint radio and IR constraints are required to verify this hypothesis.

\subsection{Constraining the SF contribution to the radio emission in the overall QSO population}\label{sec:red_emis_overall}
\begin{figure}
    \centering
    \includegraphics[width=3.3in]{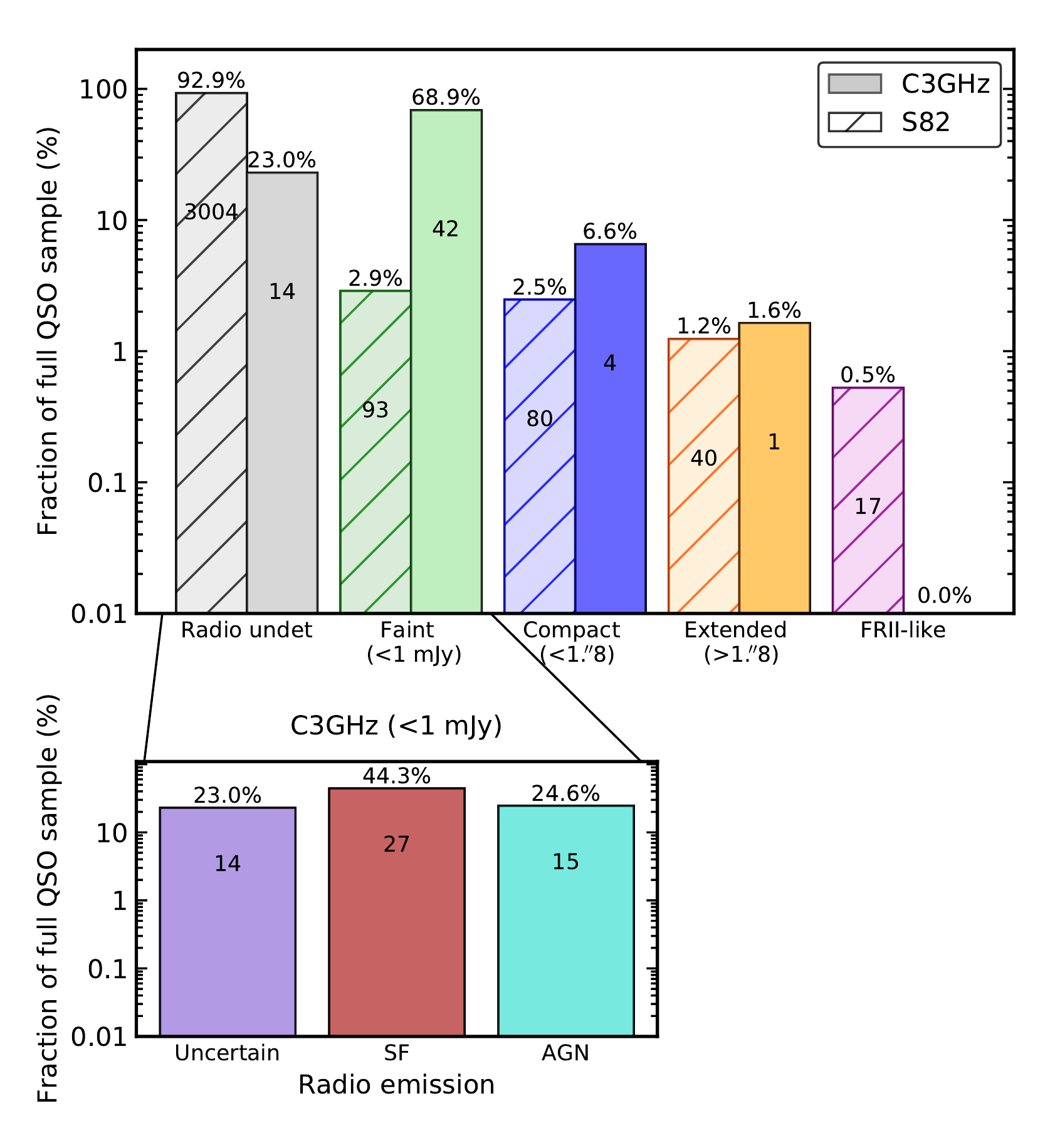}
    \caption{Origin of the radio emission as a fraction of all of the QSOs in the COSMOS and Stripe 82 regions. The radio faint class contains sources that have \textit{F}\textsubscript{1.4GHz}\,$<$\,1~mJy and so would be classified as faint in S82 but can be further explored using the greater depth of C3GHz (see zoom-in plot). The radio bright (\textit{F}\textsubscript{1.4GHz}\,$>$\,1~mJy) sources are further categorised by radio morphology (see Section~\ref{sec:morph}); the majority of the radio bright QSOs have a compact morphology ($<$\,1$\farcs$8). The number of sources in each category and associated percentage of the full QSO sample are displayed on the bars; we caution that the C3GHz compact and extended categories are highly uncertain due to poor source statistics. The zoom-in plot utilises the C3GHz depth to explore the radio-undetected and faint categories, splitting them into uncertain (upper limit on SFR within the SF-dominated region), SF (radio-detected and SF dominated; see Fig.~\ref{fig:sf}) and AGN (radio-detected but not SF dominated).}
    \label{fig:rad_emis}
\end{figure}

With the additional information on the number of SF-dominated QSOs, we can give an estimate on the overall fraction of QSOs where the radio emission is either AGN dominated or SF dominated. Fig.~\ref{fig:rad_emis} displays the fraction of all the S82 and C3GHz parent QSOs within the different radio emission categories: radio-undetected, faint (\textit{F}\textsubscript{1.4GHz}\,$<$\,1~mJy), compact, extended or FRII-like. For the bright (\textit{F}\textsubscript{1.4GHz}\,$>$\,1~mJy) source morphologies we used the S82 resolution (1$\farcs$8) for all sources. Of the radio-detected QSOs, the faint and compact categories have the highest detection fraction. The zoom in plot utilises the C3GHz depth to further explore the undetected and faint category at the S82 depth, splitting these sources into uncertain, SF dominated or AGN dominated. The uncertain category is defined by sources with an upper limit on the SFR, the SF-dominated sources have a radio luminosity within a factor of 3 of the radio emission expected from the SFR (see Fig.~\ref{fig:rad_emis}) and the AGN-dominated sources are those not uncertain nor SF dominated. The resulting estimate for the percentage of QSOs that have radio emission dominated by SF is $\approx$\,44\% (27/61), compared to $\approx$\,25\% (15/61) for AGN-dominated faint sources, $\approx$\,8\% (5/61) for AGN-dominated bright sources and $\approx$\,23\% (14/61) that are still uncertain. Including sources that are classified as uncertain, this gives a range of 44--67\% for QSOs that have radio emission dominated by SF, and 33--56\% for QSOs that have radio emission dominated by AGN processes. 

While deeper surveys such as C3GHz are limited by source statistics, this gives an indication of the capabilities of future sensitive, large-area radio surveys to constrain the origin of the radio emission in QSOs. Upcoming analysis of a sample of 90 colour-selected QSOs with FIR continuum measurements from the Atacama Large Millimetre Array (ALMA) will help us to cleanly measure the SF contribution.
\section{Conclusions}
Using high-resolution and deep radio data of SDSS DR14 QSOs, we extended the results from \cite{klindt} to further explore differences in the radio properties of red QSOs compared to typical QSOs at 0.2\,$<$\,$z$\,$<$\,2.4. With the VLA S82 and C3GHz radio data explored in our study, we probe down to host-galaxy scales and constrain the star-formation contribution to the radio emission. Our main findings are:

\begin{itemize}
    \item \textbf{rQSOs show an enhancement in radio emission around the radio-quiet threshold (see Fig.~\ref{fig:det}, \ref{fig:radio_quiet_sf}):} we confirm that rQSOs have a higher radio-detection fraction compared to cQSOs in S82, C3GHz and FIRST within uncertainties. We find that down to the S82 flux limits, this enhancement is broadly constant at a factor of $\approx$\,3 (see Section~\ref{sec:lum}). Splitting the S82 and C3GHz radio-detected sources into four contiguous $\mathcal{R}$ bins, we find that the enhancement arises from sources within $-$5\,$<$\,$\mathcal{R}$\,$<$\,$-$3.4, with no significant enhancement at higher or lower $\mathcal{R}$ values. From median stacking of the undetected S82 rQSOs and cQSOs, we find relatively higher flux values for the rQSOs when compared to cQSOs (factor $\sim$\,1.3, see Section~\ref{sec:stack}).
    \item \textbf{rQSOs show differences in their radio morphologies compared to cQSOs down to host-galaxy scales (see Fig.~\ref{fig:morph}, \ref{fig:ext}):} we find that rQSOs show an enhancement in the compact and faint radio morphologies compared to cQSOs, with a fractional difference of around 2--4. We also find an enhancement in the extended group for rQSOs (1$\farcs$8--5$''$; 16--43~kpc at $z$\,$=$\,1.5), which is not seen at the FIRST resolution, indicating we are probing scales on which these differences occur. Additionally, none of the extended rQSOs are also extended at the FIRST resolution, compared to 75\% for the cQSOs, which suggests a factor $\approx$\,10 times more rQSOs are extended on scales of 1$\farcs$8--5$''$ than cQSOs (see Section~\ref{sec:morph}).
    \item \textbf{Overall we find that 33--56\% of the DR14 QSOs have radio emission dominated by AGN processes (see Fig.~\ref{fig:sf}, \ref{fig:rad_emis}):} these ranges are constrained from the C3GHz data where we define a source as being SF dominated if its radio luminosity lies within a factor of three of that expected from SF constrained FIR data and AGN dominated otherwise. This resulted in 44--67\% of C3GHz parent QSOs classified as having their radio emission dominated by SF, compared to 33--56\% that are AGN dominated. These ranges are due to $\approx$\,23\% of QSOs still classified as uncertain (upper limit on SFR). We find the QSOs with radio emission dominated by SF emerge about one order of magnitude below the radio-quiet threshold of $\mathcal{R}$\,$=$\,$-$4.2 (see Fig . \ref{fig:radio_loud_sf}). Splitting the radio emission from the C3GHz QSOs into SF dominated and AGN dominated confirms that the decrease in radio enhancement seen at the extremely radio-quiet end is likely driven by SF (see Fig.~\ref{fig:radio_quiet_sf}). However, future deep radio and rest-frame FIR observations are needed to more directly explore the uncertain category (see Section~\ref{sec:red_emis}).
\end{itemize}

These fundamental differences in the radio properties of red QSOs predominantly arise in sources with $-$5\,$<$\,$\mathcal{R}$\,$<$\,$-$3.4 and are likely driven by differences in the radio-AGN activity between red and typical QSOs. These results provide further evidence for red QSOs representing a young phase in galaxy evolution.

In future work we will use ALMA data to further constrain the star-formation properties of red QSOs and to explore how they differ to typical QSOs. We will also use even higher resolution e-MERLIN radio data to have a more comprehensive understanding of the scale of radio emission in red QSOs down to a few kpc scales, in addition to high-frequency JVLA and 5-band low-frequency GMRT data to search for differences in the radio spectral properties and SEDs of red QSOs.
\section{Acknowledgements}
We thank the referee, Eliat Glikman, for her positive and constructive comments. We acknowledge a quota studentship funded by the Science and Technology Facility Council (VAF), the Faculty of Science Durham Doctoral Scholarship (LK), the Science and Technology Facilities Council (DMA, DJR, through grant codes ST/P000541/1 and ST/T000244/1).

Funding for the Sloan Digital Sky Survey has been provided by the Alfred P. Sloan Foundation, the U.S. Department of Energy Office of Science, and the Participating Institutions. SDSS-IV acknowledges support and resources from the Center for High-Performance Computing at the University of Utah. The SDSS web site is www.sdss.org. SDSS-IV, the primary data set used in this analysis, is managed by the Astrophysical Research Consortium for the Participating Institutions of the SDSS Collaboration including the Brazilian Participation Group, the Carnegie Institution for Science, Carnegie Mellon University, the Chilean Participation Group, the French Participation Group, Harvard-Smithsonian Center for Astrophysics, Instituto de Astrof\'isica de Canarias, The Johns Hopkins University, Kavli Institute for the Physics and Mathematics of the Universe (IPMU) / 
University of Tokyo, the Korean Participation Group, Lawrence Berkeley National Laboratory, Leibniz Institut f\"ur Astrophysik Potsdam (AIP), Max-Planck-Institut f\"ur Astronomie (MPIA Heidelberg), Max-Planck-Institut f\"ur Astrophysik (MPA Garching), Max-Planck-Institut f\"ur Extraterrestrische Physik (MPE), National Astronomical Observatories of China, New Mexico State University, New York University, University of Notre Dame, Observat\'ario Nacional / MCTI, The Ohio State University, Pennsylvania State University, Shanghai Astronomical Observatory, United Kingdom Participation Group, Universidad Nacional Aut\'onoma de M\'exico, University of Arizona, University of Colorado Boulder, University of Oxford, University of Portsmouth, University of Utah, University of Virginia, University of Washington, University of Wisconsin, Vanderbilt University, and Yale University.

This publication makes use of data products from the Wide-
field Infrared Survey Explorer, which is a joint project of the University of California, Los Angeles, and the Jet Propulsion Laboratory/California Institute of Technology, funded by the National Aeronautics and Space Administration.

The National Radio Astronomy Observatory is a facility of the
National Science Foundation operated under cooperative agreement
by Associated Universities, Inc.
\bibliography{bib}
\bibliographystyle{mnras}
\appendix
\section{\lowercase{e}BOSS sources in S82}\label{sec:bias}
For the final sample used in the analyses throughout this paper, the eBOSS-targeted QSOs were removed (see Section~\ref{sec:sample}). This was due to the discrepancy in targeting of the two separate regions of Stripe~82 (often referred to as East; RA\,$\lesssim$\,36$\degree$ and West; RA\,$\gtrsim$\,330$\degree$), with eBOSS-targeted QSOs only present in the East field, which resulted in vastly different source densities (East: 159.5~deg$^{-2}$, West: 61.4~deg$^{-2}$). After removing these sources, this reduced the source density of the East field to 58.6~deg$^{-2}$, comparable to the West field and so the two regions could be combined for our analyses.

To check that the removal of these sources did not affect the main results, our analyses were repeated using the final sample, plus the 50 eBOSS-targeted QSOs from the East field. When including the eBOSS QSOs, the enhancement in the radio-detection fraction of rQSOs compared to cQSOs was still significant at a factor $\approx$\,2.6 times higher for the rQSOs (compared to 3.3, see Section~\ref{sec:lum}). The stacked radio-undetected colour-selected S82 rQSOs also appear brighter in the radio than the cQSOs for all redshifts. For the morphology analysis, there was still an enhancement in the faint, compact and extended categories for rQSOs of 4.0$^{+1.0}_{-0.7}$, 1.9$^{+0.5}_{-0.4}$ and 2.2$^{+1.1}_{-0.6}$ respectively (compared to 4.2$^{+1.3}_{-0.9}$, 2.9$^{+0.9}_{-0.6}$ and 2.2$^{+1.6}_{-0.8}$ for the faint, compact and extended categories, respectively, see Section~\ref{sec:morph}).

Removing or including the eBOSS-targeted QSOs therefore gives qualitatively the same results, which shows that our QSO selection is not driven by the SDSS targeting approach.

\section{Spurious matches from larger radius matching}
\begin{figure*}
    \begin{center}
    \includegraphics[width=4.5in]{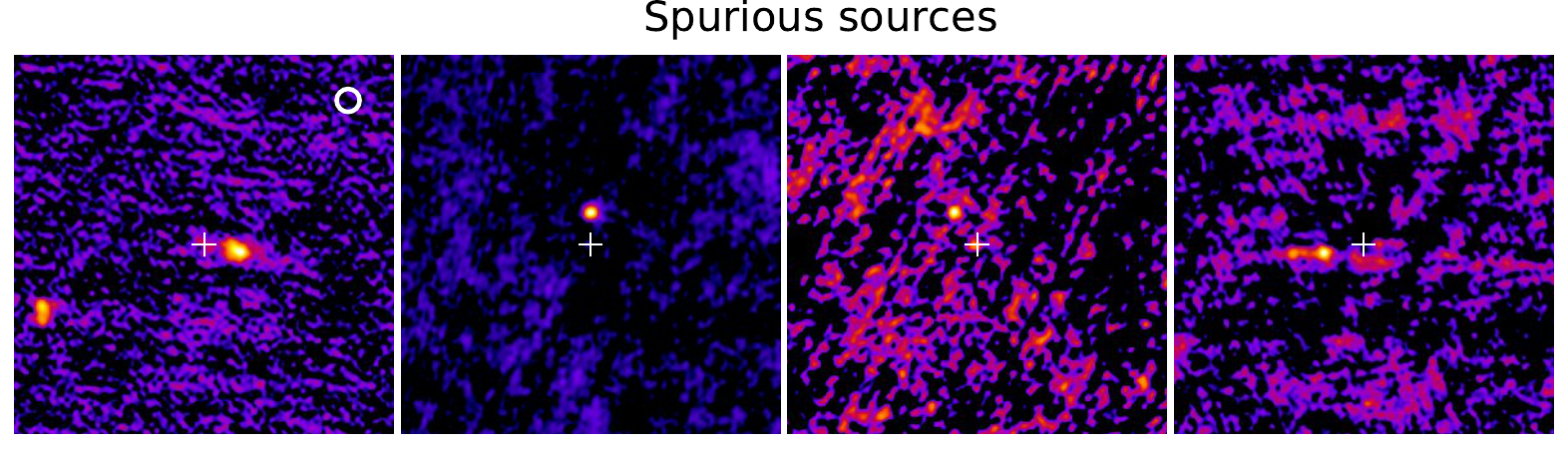}
    \hspace{0.5cm}
    \includegraphics[width=1.14in]{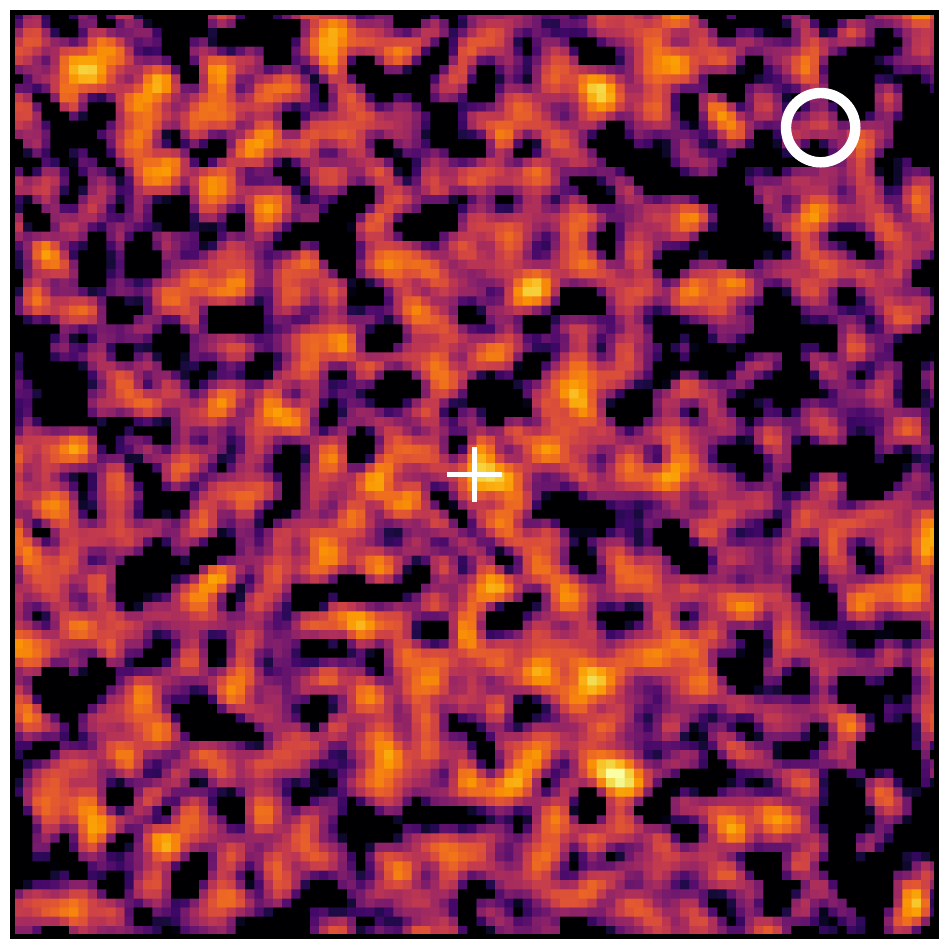}
    \end{center}
    \caption{Thumbnails (1$'$\,$\times$\,1$'$) of potential S82 spurious matches (left), and a 20$''$\,$\times$\,20$''$ thumbnail of the potential C3GHz spurious match (right) found using the 10$''$ search radius; these matches are not included in the final sample. The white cross indicates the optical QSO position and the circle on the first thumbnail displays the survey beam size (S82: 1$\farcs$8; C3GHz: 0$\farcs$75).}
    \label{fig:spurious}
\end{figure*}

Fig.~\ref{fig:spurious} displays the potential spurious matches that were not included in the final sample after matching with the larger 10$''$ radius to search for FRII systems with a weak radio core. The thumbnails on the left show the four S82 sources and the thumbnails on the right show the one C3GHz source. In each image a radio source is seen that is off-centre to the QSO position and does not appear to be associated with the QSO. For each source, a larger scale image was inspected to search for additional radio lobes that could be associated with a faint core but none were found.

\section{Comparison with FIRST}\label{sec:first}
\begin{figure}
    \centering
    \includegraphics[width=3.3in]{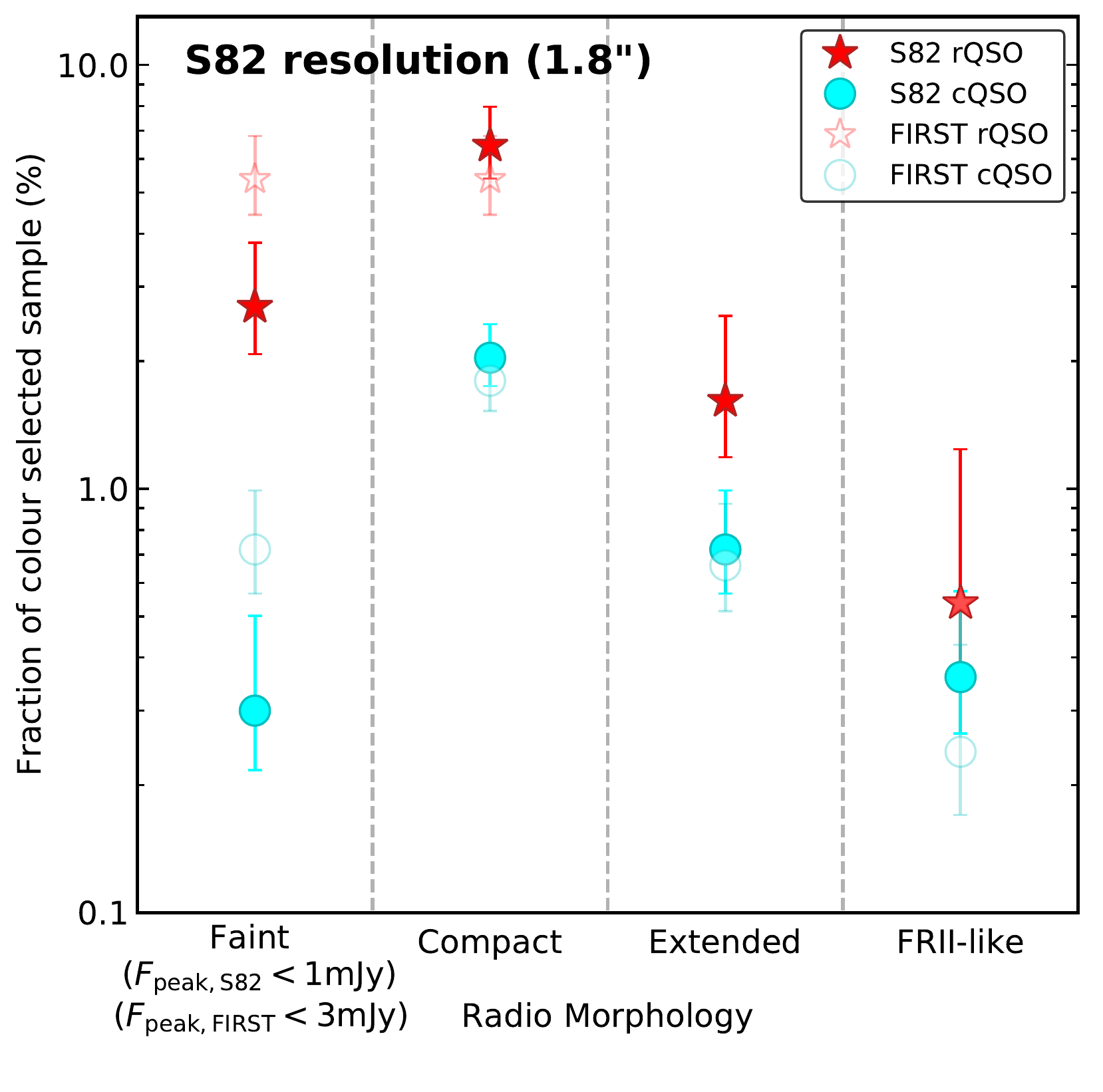}
    \caption{Radio morphology fractions of rQSOs and cQSOs for the S82-FIRST matched subset (99 QSOs) at the sensitivity limit of FIRST. This plot indicates the impact of spatial resolution on our previous results with the FIRST data \protect\citep{klindt}. Filled and open markers indicate the morphologies obtained using the S82 ($1\farcs8$ resolution) and FIRST (5$''$ resolution) data, respectively. The red arrow gives a $3\sigma$ upper limit for the FIRST extended rQSOs, due to that category containing no sources. The enhancement in the faint category for the FIRST data decreases when using the increased depth of the S82 data since we are now able to morphologically classify radio-detected QSOs with 1.4~GHz flux densities of 1--3~mJy. Similarly, the extended category shows an enhancement for rQSOs in the S82 data due to the reclassification of FIRST compact sources in the higher resolution data (see Fig.~\ref{fig:ext}).}
    \label{fig:morph_first}
\end{figure}

To explore the effect of using higher resolution, deeper radio data to classify morphologies, our sample of S82 QSOs was cross-matched with the FIRST catalogue using a 10$''$ search radius (false association rate of $\sim$\,0.3\%). This resulted in a sample of 99 QSOs (42 rQSOs and 57 cQSOs) down to the sensitivity limit of FIRST, but at 1$\farcs$8 resolution and higher SNR data given the greater sensitivity of S82. This analysis provides a more direct comparison to \cite{klindt}. 

\begin{figure}
    \centering
    \includegraphics[width=1.5in]{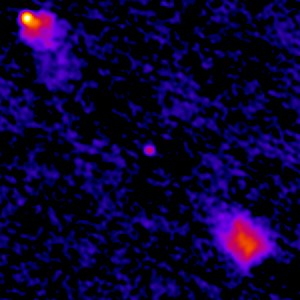}
    \hspace{1mm}
    \includegraphics[width=1.5in]{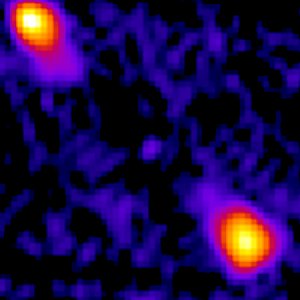}
    \caption{Thumbnails (1$\farcm$5\,$\times$\,1$\farcm$5) of the one QSO classified as FRII-like in the S82 data not detected by FIRST at the SDSS QSO position due to the faint radio core. The left image displays the source in the S82 data and the right image displays the source in the FIRST data.}
    \label{fig:fr2_first}
\end{figure}

All extended and compact S82 QSOs were detected by FIRST, and only one FRII source had no FIRST detection at the optical QSO position ($z$\,$=$\,0.6 rQSO), due to the faint core (see Fig.~\ref{fig:fr2_first}). The remaining 65 S82 sources not detected by FIRST have S82 1.4~GHz fluxes that fall below 1~mJy. A total of 28 out of 99 of the matched sources were reclassified to a different morphology type: there were no faint or compact sources that were reclassified to an FRII-like morphology and no faint sources that were reclassified as extended with the higher resolution, deeper data; all of the faint sources were classified as compact.

Fig.~\ref{fig:morph_first} displays the morphology fractions for the S82-FIRST sample; the filled markers indicate the morphology classifications obtained from the S82 data (at 1$\farcs$8 resolution), compared to the empty markers which are from the FIRST data (5$''$ resolution). It is important to note that sources classified as faint in FIRST have \textit{F}\textsubscript{peak,FIRST}\,$<$\,3~mJy, whereas for S82 they require \textit{F}\textsubscript{peak,S82}\,$<$\,1~mJy. Similarly, compact in FIRST refers to sources with no extended emission beyond the 5$''$ beam size compared to 1$\farcs$8 in S82. In the FIRST data rQSOs have a preference for faint morphologies (in agreement with \citealt{klindt}); however, due to the greater depth of S82 this difference subsides and boosts the red compact fraction. Due to the higher resolution of the S82 data, 6 rQSOs were reclassified from compact to extended. Since there were no extended rQSOs in the FIRST data, and no faint S82 sources that were reclassified as extended, this indicates that the enhancement in the extended category seen in Section~\ref{sec:morph} is driven mainly by the effect of higher resolution rather than the increased depth of the radio data. 

\label{lastpage}
\end{document}